\documentclass[lettersize,journal]{IEEEtran}
\usepackage{amsmath,amsfonts}
\usepackage{algorithm}
\usepackage{array}
\usepackage[caption=false,font=footnotesize,labelfont=rm,textfont=rm]{subfig}
\usepackage{textcomp}
\usepackage{stfloats}
\usepackage{float}
\usepackage{url}
\usepackage{verbatim}
\usepackage{graphicx}
\usepackage{bm}
\usepackage{array}
\usepackage{booktabs}
\usepackage{threeparttable}
\usepackage{multicol}  
\usepackage{multirow}  
\usepackage{amsthm, amssymb}
\usepackage{mathrsfs}
\usepackage[top=2cm, bottom=2cm, left=2cm, right=2cm]{geometry}
\usepackage{algorithmicx}
\usepackage{algpseudocode}
\usepackage{cite} 
\usepackage{colortbl}
\usepackage[font=normalsize, textfont=sf, justification=centering]{caption}
\usepackage{booktabs}
\definecolor{mygray}{gray}{.9}
\floatname{algorithm}{Algorithm}
\renewcommand{\algorithmicrequire}{\textbf{Initialize:}}

\newcommand{\tabincell}[2]{\begin{tabular}{@{}#1@{}}#2\end{tabular}}  
\begin{document}
	\title{Latency Minimization for IRS-enhanced Wideband MEC Networks with Practical Reflection Model }
	\author{Nana Li, Wanming Hao,~\IEEEmembership{Senior Member,~IEEE,} Xingwang Li,~\IEEEmembership{Senior Member,~IEEE,} Zhengyu Zhu,~\IEEEmembership{Senior Member,~IEEE,}
		Zhiqing Tang~\IEEEmembership{Member,~IEEE,} Shouyi Yang
		\thanks{N. Li, W. Hao, Z. Zhu and S. Yang are with the School of Electrical and Information Engineering, Zhengzhou University, Zhengzhou 450001, China. (e-mail: nnli@gs.zzu.edu.cn, iewmhao@zzu.edu.cn, iezyzhu@zzu.edu.cn, iesyyang@zzu.edu.cn)}
			\thanks{X. Li is with the School of Physics and Electronic Information Engineering, Henan Polytechnic University, Jiaozuo, China. (email: lixingwang@hpu.edu.cn)}
			\thanks{Z. Tang is with the School of Cyber Science and Engineering, Zhengzhou University, Zhengzhou 450002, China. (e-mail: iezqtang@zzu.edu.cn)}
		}
	\maketitle
	
	\begin{abstract}
		Intelligent reflecting surface (IRS) has been considered as an efficient way to boost the computation capability of mobile edge computing (MEC) system, especially when the communication links is blocked or the communication signal is weak. However, most existing works are restricted to narrow-band channel and ideal IRS reflection model, which is not practical and may lead to significant performance degradation in realistic systems. To further exploit the benefits of IRS in MEC system, we consider an IRS-enhanced wideband MEC system with practical IRS reflection model. With the aim of minimizing the weighted latency of all devices, the offloading data volume, edge computing resource, BS's receiving vector, and IRS passive beamforming are jointly optimized. Since the formulated problem is non-convex, we employ the block coordinate descent (BCD) technique to decouple it into two subproblems for alternatively optimizing  computing and communication settings.  The effectiveness and convergence of the proposed algorithm are validate via numerical analyses. In addition, simulation results demonstrate that the proposed algorithm can achieve lower latency compared to that based on the ideal IRS reflection model, which confirms the necessary of considering practical model when designing an IRS-enhanced wideband MEC system.
	\end{abstract}
	

	\section{Introduction}
	\IEEEPARstart{W}{ith} the rapid development of communication technology, a large number of compute-intensive and delay-sensitive applications have emerged, such as virtual reality, autonomous driving, and digital twin \cite{ref1,ref2,ref3}. Because of the restricted computing capabilities, it is challenging for these wireless devices (WDs) to fulfill the low latency and high reliability computing needs of the applications. Mobile edge computing (MEC) \cite{ref4,ref5} technology can help reduce the computational load of WDs and enhance computing performance. However, the benefits of MEC cannot be fully exploited when the communication links between WDs and base station (BS) are obstructed \cite{ref6}.  To handle this issue, the intelligent reflective surface (IRS) can be utilized to build an IRS-enhanced MEC system.

	IRS can be deployed easily to improve communication environment for transmitters and receivers \cite{ref7,ref8}. Nonetheless, how to design effective IRS reflection coefficients is still worth further research. Existing researches on IRS-enhanced wireless communication systems are mostly limited to narrowband channels and have operated under the assumption of an ideal reflection model. This model induces a constant amplitude but varying phase shift response to incident signals. Unfortunately, achieving such an ideal IRS is challenging for current hardware circuit technology \cite{ref9}. Consequently, existing designs based on this ideal model may lead to significant performance degradation in realistic systems because of the huge difference between the practical hardware circuit response and the ideal one \cite{ref10,ref11,ref12,ref13}. Therefore, in this paper, we aim to jointly optimize computing and communication resource in IRS-enhanced wideband MEC system with practical IRS reflection model. 
	
	\subsection{Prior Works}
	\subsubsection{MEC system} Depending on the interdependence and partitionability amongs tasks, the computational tasks of devices can be categorized into two typical computation offloading modes: binary offloading and partial offloading \cite{ref5}. Specifically, in binary offloading mode, the computing tasks as a whole are required to be either executed locally or offloaded to the edge server for remote processing. In contrast, for the partial offloading mode, the tasks can be divided into two parts, with one part executed locally and the other part offloaded to the edge server \cite{ref15}. Both of these computation offloading modes have attacted a lot of research interest in jointly optimizing computing and communication settings to minimize computation latency \cite{ref16,ref17,ref18}, energy consumption \cite{ref19,ref20}, and maximize energy efficiency \cite{ref21}, computation sum-rate \cite{ref22,ref23}. However, when the communication links are blocked or the received signal is weak, the data processing capability of MEC may be influenced, resulting in limited utilization efficiency of edge computation resource.
	
	\subsubsection{IRS-enhanced MEC systems}
	To enhance the task offloading efficiency, only a few works have considered introducing IRSs into MEC systems with the objectives of latency minimization \cite{ref24}, computing rate maximization \cite{ref25,ref26,ref27}, and energy consumption minimization \cite{ref28,ref29,ref30}. In \cite{ref31} and \cite{ref24}, Bai et al. first provided an overview of the IRS-enhanced MEC and then minimized the overall latency of all WDs by optimizing WDs' offloading data volume, edge computing resource, IRS phase shift and BS's receiving vector. In \cite{ref26} and \cite{ref27}, the authors formulated a computation rate maximization problem under the limited energy budget, while the latency imposed by edge computing is ignored by assuming sufficient edge computation resources. To maximize the total computing data bits of WDs, the authors in \cite{ref25} further provided a deep learning-based method to reduce the computational complexity of traditional optimization algorithms.
	In \cite{ref28,ref29,ref30}, the authors studied the energy consumption minimization of the IRS-enhanced MEC system by jointly optimizing the computing resource of edge server and IRS passive beamforming.

	\subsubsection{Motivations and contributions}
	MEC and IRS are two candidate technologies for future wireless communication. The research on IRS-enhanced MEC system is currently in its early stage. To support different applications, the computing and communication resource must be carefully allocated to fulfill different requirements. The latency is utmost importance for improving the performance of MEC systems. Beside, since WDs' computing offloading mode and IRS reflection optimization interact with each other, the optimization method under ideal model proposed in \cite{ref24,ref25,ref28,ref29,ref27,ref30,ref31,ref26} may not suitable for the non-ideal optimization design. In our previous work \cite{ref32}, we also investigated the resource allocation in IRS-MEC system with non-ideal reflection model. Nevertheless, the IRS-MEC systems mentioned above are limited to narrow-band channels, and for the wideband communication, the response characteristics of IRS also vary with the incident signal frequency.
	
	Therefore, to exploit the full potential of the IRS-enhanced wideband MEC system, in this paper, we focus on minimizing the weighted latency using a practical IRS reflection model. The main contributions are summarized as follows.
	\begin{itemize}
		\item[•] \emph{IRS-enhanced wideband MEC system with practical reflection model:} Different from the existing works which focus on the narrow-band system and ideal IRS reflection model, we aim to improve the computation performance of wideband MEC by exploiting the full benefits of IRS based on the practical reflection model. A weighted latency minimization problem is formulated for multiple devices scenario, which jointly optimizes the offloading data volume, edge computing resources, BS's receiving vectors, and BPS at IRS, subject to both constraints of limited edge computing capability and non-convex BPS matrix. The coupling effect among multiple variables makes it challenging to find a global optimal solution to this problem. Therefore, a block coordinate descent (BCD) technique-based method is proposed to decouple the original problem into two subproblems, and then the computing and communication settings are designed alternately. 		
		
		\item[•] \emph{Computing design:} Fixing the communication setting, the offloading data volume and edge computing resource can be decoupled by again using BCD technique. Given edge computing resources, the optimal offloading data volume can be obtained by making the local computing latency equal to that of edge computing. Upon obtaining the semi-closed solution of offloading data volume, the subproblem is transformed to a convex problem, and then relying on the Karush–Kuhn–Tucker (KKT) conditions and classic bisection search method, the optimal solution of edge computing resources can be found. 
		
		\item[•] \emph{Communication design:} Based obtained computing setting, it is still challenging to solve this subproblem directly due to the non-convex sum-of-ratio form of objective function. By introducing auxiliary variables and levaraging Lagrangian dual reformulation (LDR) technique, this subproblem is transformed into an equivalent form, and then an two stage method is proposed to find the optimal solution. In each iteration, the auxiliary variables and lagrangian multiplier are updated using a Newton-like method, while the solutions of BS's receiving vectors and IRS phase shifts are obtained by leveraging the equivalence relationship between sum-rate maximization and mean square error (MSE) minimization.
		
		\item[•] \emph{Simulation analysis:} Finally, numerical results verify the convergence of the proposed algorithm. Moreover, we evaluate the proposed algorithm by extensive simulation analysis, which confirms the effectiveness of the considered algorithm with practical model compared to that with the ideal one.
	\end{itemize} 
	
	\section{System Model and Problem Formulation}
	\begin{figure}[htbp]
		\begin{minipage}[t]{0.42\textwidth}
			\centering
			\includegraphics[width=0.9\textwidth]{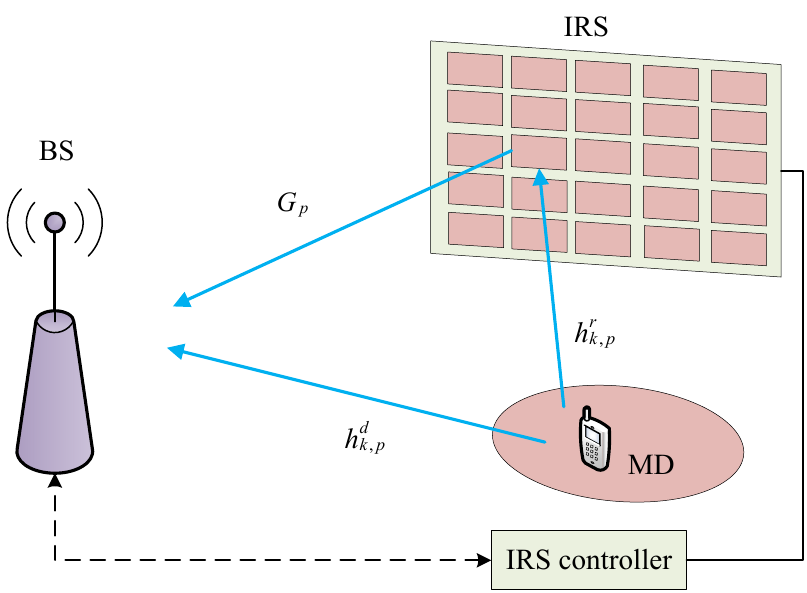}
			\caption{IRS-enhanced wideband MEC system model.}
			\label{fig1}
		\end{minipage}
	\end{figure}
	\subsection{System Model}
	As shown in Fig. \ref{fig1}, we consider a wideband multi-user orthogonal frequency division multiplexing (MU-OFDM) system with $P$ subcarriers. $K$ single-antenna WDs offload part of computing data to the BS equipped with edge servers with the help of IRS. $M$ and $N$ represent the number of BS antennas and IRS reflection elements, respectively. Denote $ \mathcal{P}=\left\lbrace 1,...,P \right\rbrace  $, $ \mathcal{K} =\left\lbrace1,...,K \right\rbrace  $, $ \mathcal{M}=\left\lbrace1,...,M \right\rbrace  $, and $ \mathcal{N}=\left\lbrace 1,...,N\right\rbrace  $ as the set of subcarriers, users, transmission antennas, and IRS elements. The edge node and BS are deployed together and connected via a high-speed, low-latency wired fiber. Thus, we ignore the data transmit delay between BS and edge nodes\cite{ref33,ref34}. 
	
	\subsection{IRS Reflection Model}
	In \cite{ref12}, Cai et al. established a three-dimensional reflection model to accurately describe the variation of IRS amplitude and phase shift with BPS and  incident signal frequency. This model is so complicated that it may lead to the optimization design of IRS more difficult and complex. Hence, in \cite{ref35}, they further proposed a simplified model based on the actual application. When the relative bandwidth, i.e., $B/f_{c}$, is less than 5\%, the relationship between the reflection amplitudes and phase shifts of IRS elements can be viewed as a quadratic function, and the curves for different frequencies do not have obvious difference. Here, $B$ and $ f_{c} $ represent bandwidth and carrier frequency, respectively. In addition, phase shift can be approximately fitted as a straight line with respect to frequency, whose slope and intercept vary with the basic phase shift (BPS) $\theta$. Here, BPS $\theta$ is defined as the phase shift for carrier frequency $f_{c}$. Upon denoting the frequency of incident signal by $f$, the reflection amplitude $\mathrm{A} (\theta, f)$ and phase shift $ \mathrm{B}(\theta, f) $ are given by 
	\begin{equation}\label{Eq1}
		\begin{aligned}
			&\mathrm{A} (\theta, f)  =\alpha_1 \mathrm{B}^2(\theta, f)+\beta_1 \mathrm{B}(\theta, f)+\tilde{c}_1, \\
			&\mathrm{B}(\theta, f)  =\mathcal{F}_{1}(\theta) f+\mathcal{F}_{2}(\theta), \\
			&\mathcal{F}_{1}(\theta)  =\alpha_2 \sin \left(\beta_2 \theta+\tilde{c}_2\right)+\alpha_3 \sin \left(\beta_3 \theta+\tilde{c}_3\right), \\
			&\mathcal{F}_{2}(\theta)  =\alpha_4 \sin \left(\beta_4 \theta+\tilde{c}_4\right)+\alpha_5 \sin \left(\beta_5 \theta+\tilde{c}_5\right),
		\end{aligned}
	\end{equation} 
	where $ \mathcal{F}_{1}(\theta) $ represents the slope and $ \mathcal{F}_{2}(\theta) $ represents the intercept of the phase shift-frequency line, respectively. $ \left\lbrace \alpha_{i}, \beta_{i}, \tilde{c}_{i} \right\rbrace_{i=1}^{5}  $ are IRS parameters related to the specific cuicuit implementation. For instance, when the carrier frequency $f_{c}=2.4$ GHz and bandwidth $B=100$ MHz, the parameter values are shown in Table \ref{tab1}. In addition, the central frequency $f_i$ of each subcarrier is given by $ f_p \triangleq f_{\mathrm{c}}+\left(p-\frac{P+1}{2}\right) \frac{B}{P} $, $ \forall p \in \mathcal{P} $.

	\begin{table}[t]\small
		\centering
		\caption{\quad Parameters Values for Non-ideal IRS Models}
		\begin{tabular}{cccccc}
			\toprule[1.2pt]
			& i=1 & i=2 & i=3 & i=4 & i=5 \\
			\midrule[0.8pt]
			$ \alpha_{i} $ & 0.06&11.27&10.88&89.64&26.11\\
			$ \beta_{i} $ & 0.02&0.008996&0.9799&0.01268&0.9798\\
			$ \tilde{c}_{i} $ & 0.5736&-1.897&-1.471&0.2899 &1.673 \\
			\bottomrule[1.2pt]
		\end{tabular}
		\label{tab1}
	\end{table}
	
	\subsection{Communication Model}	
	Let $ \mathbf{h}_{k, p}^{\mathrm{d}} \in \mathbb{C}^{M \times 1} $, $ \mathbf{G}_{p} \in \mathbb{C}^{M \times N}$, and $ \mathbf{h}_{k,p}^{r} \in \mathbb{C}^{N \times 1}$ denote the baseband equivalent channels from the $k$-th WD to BS, from IRS to BS, and from the $k$-th WD to IRS on the $p$-th subcarrier. Let $\boldsymbol{\Phi}_{p}=\operatorname{diag}\left( \phi_{p,1},...,\phi_{p,N} \right) $ denote the reflection matrix of IRS on the $p$-th subcarrier. Here, $\phi_{p, n}$ is the reflection coefficient of the $n$-th IRS element on the $p$-th subcarrier, which satisfies the following functional relationships 
	\begin{equation}\label{Eq3}
		\left| \phi_{p,n} \right|=\mathrm{A}\left( \theta_{n}, f_{p} \right), \angle \phi_{p,n}=\mathrm{B}\left( \theta_{n}, f_{p} \right), \forall p, n,  
	\end{equation}
	where $ \theta_{n} $ is the BPS of the $n$-th IRS element and $ \boldsymbol{\theta}=\left[ \theta_{1},...,\theta_{N} \right]^{T} $ denotes the BPS vector. Let $ \mathbf{s}_{p} \triangleq \left[s_{1,p},...,s_{K,p} \right]^{T} \in \mathbb{C}^{K} $ represent the transmission signal of the $K$ WDs on the $p$-th subcarrier, and satisfy $ \mathbb{E}\left\lbrace \mathbf{s}_{p}\mathbf{s}_{p}^{H} \right\rbrace =\mathbf{I}_{K} $, $\forall p \in \mathcal{P} $. Upon denoting the transmit power of WD by $\widetilde{p}$, the received signal $ \mathbf{y}_{p} \in \mathbb{C}^{M \times 1}$ at BS on the $p$-th subcarrier can be formulated as
	\begin{equation}\label{Eq4}
		\mathbf{y}_{p}=\sqrt{\widetilde{p}}\sum_{k=1}^{K}\left( \mathbf{h}_{k, p}^{\mathrm{d}}+ \mathbf{G}_{p}\boldsymbol{\Phi}_{p} \mathbf{h}_{k,p}^{r} \right)s_{k,p}+\mathbf{n}_{p}, 
	\end{equation}
	where $ \mathbf{n}_{p} \triangleq \left[n_{p,1},...,n_{p,M} \right]^{T}  $ is the noise vector, which satisfies $ n_{p, m} \sim \mathcal{C}\mathcal{N} \left( 0,\sigma^{2} \right) $, $ \forall p \in \mathcal{P} $, $m \in \mathcal{M}$.  Upon denoting the BS's receiving vectors for the $k$-th WD on the $p$-th subcarrier by $\mathbf{u}_{k,p}$, the recovered signal for the $k$-th WD on the $p$-th subcarrier is given by
	\begin{equation}\label{Eq5}
		\hat{y}_{k, p}\!=\!\mathbf{u}_{k,p}^{H} \left( \sqrt{\widetilde{p}} \sum_{j=1}^{K}\left( \mathbf{h}_{j, p}^{\mathrm{d}}\!+\!\mathbf{G}_{p}\boldsymbol{\Phi}_{p} \mathbf{h}_{j,p}^{r} \right) s_{j,p}\!+\!\mathbf{n}_{p}\right).
	\end{equation} 
	Then, the SINR of the $k$-th WD on the $p$-th subcarrier can be expressed as
	\begin{equation}\label{Eq6}
		\gamma_{k,p}\left(\mathbf{u}_{k,p}, \boldsymbol{\Phi}_{p}\right)\!=\!\frac{\widetilde{p}\left|\mathbf{u}_{k,p}^H\left(\mathbf{h}_{k,p}^{d}+\mathbf{G}_{p} \boldsymbol{\Phi}_{p} \mathbf{h}_{k,p}^{r}\right)\right|^2}{\widetilde{p}\!\sum_{j=1, j \neq k}^K\!\left|\mathbf{u}_{k,p}^H\left(\mathbf{h}_{j,p}^{d}\!+\!\mathbf{G}_{p} \boldsymbol{\Phi}_{p} \mathbf{h}_{j,p}^{r}\right)\right|^2\!\!\!+\!\sigma^2}.
	\end{equation}
	Accordingly, we can obtain the offloading rate for the $ k $-th WD on the $ p $-th subcarrier as
	\begin{equation}\label{Eq7}
		R_{k,p}\left(\mathbf{u}_{k,p},\boldsymbol{\Phi}_{p}\right)   =B \log_{2} \left[ 1+ \gamma_{k,p}\left(\mathbf{u}_{k,p}, \boldsymbol{\Phi}_{p}\right) \right], 
	\end{equation}
	The total transmission rate for the $ k $-th user is given by 
	\begin{equation}\label{Eq8}
		R_k(\mathbf{u}_{k}, \boldsymbol{\theta})=\sum_{p=1}^{P} R_{k,p}\left(\mathbf{u}_{k,p},\boldsymbol{\Phi}_{p}\right), \forall k, p, 
	\end{equation}
	where $ \mathbf{u}_{k}=[\mathbf{u}_{k,1}^{T},...,\mathbf{u}_{k,P}^{T}]^{T} $ represent the BS's receiving vector for the $k$-th WD.
		
	  \subsection{Computing Model}
		We consider a partial offloading application model where the computing data is arbitrarily divisible \cite{ref36}. WD can offload a part of computing data to the edge server for remote processing, and the other part for local computing.
		\begin{itemize}
			\item{Local computing:} Let $D_{k}$, $d_{k}$, and $c_{k}$ denote the total computation data volume, offloading data volume, and computational complexity of the computation data. Upon denoting the computing power of the $k$-th WD by $F_{k}^{l}$, the local computing latency is given by $ T_{k}^{l}=(D_{k}-d_{k})c_{k}/F_{k}^{l} $.
			
			\item{Edge processing:} Denote $ F_{\rm{total}}^{e} $ and $ F_{k}^{e} $ as the total computing capacity of edge server and the edge computing resource allocated to the $k$-th WD, respectively, satisfying $ \sum_{k=1}^{K} F_{k}^{e} \leq F_{\rm{total}}^{e} $. The latency usually consists of three parts: 1) offloading latency for data transmission; 2) computing latency for processing the offloaded data on the edge server; 3) return latency for transmitting the computing result to WD. Generally, due to the small data volume of computing result, the return latency of the last part is usually ignored \cite{ref32,ref38}. Thus, the latency introduced by edge computing is expressed as $ T_{k}^{e}\left(\mathbf{u}_{k} , \boldsymbol{\theta}, d_{k}, F_{k}^{e} \right)={d_{k}}/{R_{k}\left(  \mathbf{u}_{k}, \boldsymbol{\theta} \right) }+{d_{k}c_{k}}/{F_{k}^{e}}  $.
		\end{itemize}
		
		The overall latency of the $k$-th WD can be expressed as 
		\begin{equation}\label{Eq9}
			\begin{aligned}
				T_{k}\!\left( \mathbf{u}_{k},\!\boldsymbol{\theta},\! d_{k}, \! F_{k}^{e} \right)  &\! = \!\max \left\lbrace T_{k}^{l}\left( d_{k} \right), T_{k}^{e}\left(  \mathbf{u}_{k}, \boldsymbol{\theta}, d_{k}, F_{k}^{e}  \right)  \right\rbrace , \\
				&\! =\!\max\left\lbrace \! \dfrac{(\!D_{k}\!-\!d_{k}\!)c_{k}}{F_{k}^{l}}\!,\! \dfrac{d_{k}}{R_{k}\!\left( \mathbf{u}_{k},\! \boldsymbol{\theta} \right) }\!+\!\dfrac{d_{k}c_{k}}{F_{k}^{e}} \right\rbrace,
			\end{aligned}
		\end{equation}
		
		\subsection{Problem Formulation}
		In this paper, we investigate a weighted computational latency minimization problem by jointly optimizing the offloading data volume $\boldsymbol{d}=\left[ d_{1}, d_{2},...,d_{K} \right]^{T} $, edge computing resource $ \mathbf{F}^{e}=\left[ F_{1}^{e},...,F_{k}^{e} \right]  $, BS's receiving vectors $\mathbf{U}=\left[ \mathbf{u}_{1,1}^{T},...,\mathbf{u}_{K,1}^{T},...,\mathbf{u}_{K,P}^{T} \right]^{T} $, and the IRS BPS $\boldsymbol{\theta}$, which is formulated as
		\begin{subequations}\label{OptA}
			\begin{align}
				&\min _{\mathbf{U}, \boldsymbol{\theta}, \boldsymbol{d}, \mathbf{F}^e} \sum_{k=1}^K \varpi_k T_k\left(\mathbf{u}_k, \boldsymbol{\theta}, d_k, F_k^e\right)  \label{OptA1}\\
				&\text { s.t. } \left|\phi_{p, n}\right|=\mathrm{A}\left(\theta_n, f_p\right), \forall p, n,  \label{OptA2} \\
				& \qquad \angle \phi_{p, n}=\mathrm{B}\left(\theta_n, f_p\right),\forall p, n, \label{OptA3} \\
				& \qquad \theta_{n} \in \left[ -\pi,  \pi \right] , \forall n, \label{OptA4} \\
				& \qquad \left\|\mathbf{u}_{k,p} \right\|^{2} \leq 1, \forall k, p, \label{OptA5}  \\
				& \qquad d_k \in\left\{0,1, \ldots, D_k\right\}, \forall k, \label{OptA6} \\
				& \qquad \sum_{k=1}^K F_k^e \leq F_{\text {total }}^e, \label{OptA7} \\
				& \qquad F_k^e \geq 0, \forall k, \label{OptA8} 			
			\end{align}
		\end{subequations}
		where $ \varpi_k $ denotes the weight of the $k$-th WD. (\ref{OptA2}) and (\ref{OptA3}) are the amplitude and phase-shift constraints of the non-ideal IRS  model, respectively. (\ref{OptA4}) is the phase shift constraint; (\ref{OptA5}) denotes the unit-norm receiving vector constraint for the $k$-th WD on the $p$-th subcarrier; (\ref{OptA6}) indicates that the offloaded data volume should be an integer between zero and the total computation data.
		
		\emph{Remark 1: } For the optimization variables of problem (\ref{OptA}), $\left\lbrace \boldsymbol{d}, \mathbf{F}^{e} \right\rbrace $ are variables 
		related to the computing design, $\left\lbrace \mathbf{U}, \boldsymbol{\theta} \right\rbrace $ are variables related to the communication design. It is challenging to solve problem (\ref{OptA}) due to the following three aspects: 1) the segmented form of objective function; 2) the coupling effect of $\mathbf{U}$ and $\boldsymbol{\theta}$; 3) the non-convex objective function with respect to $\boldsymbol{\theta}$. To proceed, we provide a locally optimal solution by decoupling the communication and computing designs using BCD technique. Then, the segmented form of objective function can be transformed into a more tractable form. Leveraging the equivalence between sum-rate maximization and MSE minimization, the communication design subproblem is converted into a multi-variable problem, which can be solved effectively via BCD technique. To address the non-convexity of $\boldsymbol{\theta}$, we solve the BPS for each IRS reflection element separately, and then update $ \boldsymbol{\theta} $ iteratively until it converges.
		
		\section{Proposed Solution Based on BCD Technique}
		In this section, we jointly optimize the computing and communication setting relying on the BCD technique. Next, we first optimize offloading data volume and edge computing resource, while fixing the communication setting, and then optimize BS's receiving vectors and BPS with fixed computing setting. Our goal is to jointly optimize computing and communication settings until convergence.
		
		\subsection{Optimization of Offloading Data Volume and Edge Computing Resources}\label{IIIA}
		Given the BS's receiving vector $ \mathbf{U} $ and BPS $\boldsymbol{\theta}$, the original problem (\ref{OptA}) can be simplified to
		\begin{subequations}\label{OptB}
			\begin{align}
				& \min _{\boldsymbol{d}, \mathbf{F}^e} \sum_{k=1}^K \varpi_k T_k\left( d_k, F_k^e \right)  \label{OptB1}\\
				& \text { s.t. (\ref{OptA6}), (\ref{OptA7}), (\ref{OptA8})}	\label{OptB2}
			\end{align}
		\end{subequations}
		
		\emph{1) Optimization of $ \boldsymbol{d} $: } 
		Given a set of solutions for $\left\lbrace \mathbf{U}\text{,} \boldsymbol{\theta}\text{,} \mathbf{F}^{e} \right\rbrace$, the optimal offloading data volume can be determined by setting $ T_{k}^{l} ( \hat{d}_{k} ) =  T_{k}^{e} ( \hat{d}_{k} ) $, i.e.,
		\begin{equation}\label{Eq13}
			\hat{d}_k^*=\frac{D_k c_k R_k F_k^e}{F_k^e F_k^l+c_k R_k\left(F_k^e+F_k^l\right)}.
		\end{equation}
		Bear in mind that the offloading data volume must be a positive integer, so the optimal $d_k^*$ is given by
		\begin{equation}\label{Eq12}
			d_k^*=\underset{\hat{d}_k \in\left\{\left\lfloor\hat{d}_k^*\right\rfloor,\left\lceil\hat{d}_k^*\right\rceil\right\}}{\arg \min } T_k\left(\hat{d}_k\right),
		\end{equation}
		where $ \left\lfloor*\right\rfloor $ and $ \left\lceil * \right\rceil $ represent the floor and ceiling operations, respectively. 
		
		\emph{Proof:} Upon denoting the relaxation \cite{ref39} of integer value $d_{k} \in\left\{0,1, \ldots, L_{k}\right\}$ by  $\hat{d}_{k} \in\left[0, L_{k}\right]$, the overall latency of the $k$-th WD is $\hat{T}_{k}\left(\hat{d}_{k}\right) \triangleq \max \left\{T_{k}^{l}\left(\hat{d}_{k}\right), T_{k}^{e}\left(\hat{d}_{k}\right)\right\}$, which can be reformulated from (\ref{Eq9}) as the following segmented form
		\begin{equation}\label{Eq14}
			\hat{T}_{k}\left(\hat{d}_{k}\right)= \begin{cases}\frac{\left(D_{k}-\hat{d}_{k}\right) c_{k}}{F_{k}^{l}}, & 0 \leq \hat{d}_{k} \leq \frac{D_{k} c_{k} R_{k} F_{k}^{e}}{F_{k}^{e} F_{k}^{l}+c_{k} R_{k}\left(F_{k}^{l}+F_{k}^{e}\right)} \\ \frac{\hat{d}_{k}}{R_{k}}+\frac{\hat{d}_{k} c_{k}}{F_{k}^{e}}, & \frac{D_{k} c_{k} R_{k} F_{k}^{e}}{F_{k}^{e} F_{k}^{l}+c_{k} R_{k}\left(F_{k}^{l}+F_{k}^{e}\right)}<\hat{d}_{k} \leq D_{k}  \end{cases}
		\end{equation}
		From (\ref{Eq14}), we can observe that when $\hat{d}_{k}$ increases from 0 to  $ \frac{D_{k} c_{k} R_{k} F_{k}^{e}}{F_{k}^{e} F_{k}^{l}+c_{k} R_{k}\left(F_{k}^{l}+F_{k}^{e}\right)} $, the latency $\hat{T}_{k}\left(\hat{d}_{k}\right)$ decreases, and when $ \hat{d}_{k} $ increases from $ \frac{D_{k} c_{k} R_{k} F_{k}^{e}}{F_{k}^{e} F_{k}^{l}+c_{k} R_{k}\left(F_{k}^{l}+F_{k}^{e}\right)} $ to $ D_{k} $, $\hat{T}_{k}\left(\hat{d}_{k}\right)$ increases. Hence, to minimize latency, we should select $ \hat{d}_{k}=\hat{d}_{k}^{*}=\frac{D_{k} c_{k} R_{k} F_{k}^{e}}{F_{k}^{e} F_{k}^{l}+c_{k} R_{k}\left(F_{k}^{l}+F_{k}^{e}\right)} $.  Since the optimal value of $d_{k}$ should be a non-negative integer, we can obtain its optimal value after carrying out the following operation  $ d_{k}^{*}=\underset{\hat{d} \in\left\{\left\lfloor\hat{d}_{k}^{*}\right\rfloor,\left[\hat{d}_{k}^{*}\right]\right\}}{\arg \min } T_{k}\left(\hat{d}_{k}\right) $. 
		
		\emph{2) Optimization of $ \mathbf{F}^{e} $:} Next, given a set solutions of $\left\lbrace \mathbf{U}, \boldsymbol{\theta}, \boldsymbol{d} \right\rbrace $, we focus on the optimization of edge computing resource. After substituting (\ref{Eq13}) into objective (\ref{OptB1}), the corresponding computing setting subproblem can be reformulated as
		\begin{subequations}\label{OptC}
			\begin{align}
				& \min _{\mathbf{F}^e} \sum_{k=1}^K \frac{\varpi_k\left(D_k c_k^2 R_k+D_k c_k F_k^e\right)}{F_k^e F_k^l+c_k R_k\left(F_k^e+F_k^l\right)} \label{OptC1} \\
				& \text { s.t. (\ref{OptA7}), (\ref{OptA8}).}
			\end{align}
		\end{subequations}
		Next, we will prove that problem (\ref{OptC}) is a convex problem, and then a solution of $\mathbf{F}^e$ is provided by exploiting the KKT condition.
		
		Denote the second derivative of objective function (\ref{OptC1}) with respect to $F_{k}^{e}$ by $f^{''}$, which is given by
		\begin{equation}
			f^{''}=\frac{2 \varpi_k D_k c_k^3 R_k^2\left(F_k^l+c_k R_k\right)}{\left[F_k^e F_k^l+c_k R_k\left(F_k^e+F_k^l\right)\right]^3}.
		\end{equation}
		Since parameters $ \varpi_k $, $ c_k $, $ R_k $, and $F_k^l$ are all positive and we have $ D_k \geq 0 $ and $ F_k^e \geq 0 $, it can be inferred that $f^{''} \geq 0$. Besides, constraint (\ref{OptA7}) and (\ref{OptA8}) are all linear forms. Thus, we can conclude that problem (\ref{OptC}) is strictly convex, and its optimal solution satisfies the KKT condition. Specifically, the corresponding Lagrangian function is formulated as
		\begin{equation}
			\begin{aligned}
				\mathcal{L}\left(\mathbf{F}^e, \eta\right)=\sum_{k=1}^K \frac{\varpi_k\left(D_k c_k^2 R_k+D_k c_k F_k^e\right)}{c_k R_k F_k^l+\left(F_k^l+c_k R_k\right) F_k^e} \\
				+\eta\left(\sum_{k=1}^K F_k^e-F_{\rm{total }}^e\right),
			\end{aligned}
		\end{equation}
		where $\eta \geq 0$ is the Lagrangian multiplier. The optimal solution of edge resource allocation $\mathbf{F}^{e^*}$ and Lagrangian multiplier $ \eta^{*} $ should satisfy the related KKT conditions, i.e.,
		\begin{align} 
			& \frac{\partial \mathcal{L}}{\partial F_k^e}=\frac{-\varpi_k D_k c_k^3 R_k^2}{\left[c_k R_k F_k^l+\left(F_k^l+c_k R_k\right) F_k^{e *}\right]^2}+\eta^*=0, \label{Eq18}\\ 
			& \eta^*\left(\sum_{k=1}^K F_k^{e *}-F_{\text {total }}^e\right)=0, \label{Eq19} \\
			& F_k^{e *} \geq 0. \label{Eq20}
		\end{align}		
		For a given $ \eta $, the optimal $ F_{k}^{e} $ can be directly derived from (\ref{Eq18}), which is given by
		\begin{equation}\label{Eq21} 
			F_k^e=\frac{\sqrt{\frac{\varpi_k D_k c_k^3 R_k^2}{\eta}}-c_k R_k F_k^l}{F_k^l+c_k R_k}, \quad \forall k.
		\end{equation}
		To ensure $ F_{k}^{e} \geq 0 $ in (\ref{OptA8}), the numerator of formula (\ref{Eq21}) must satisfy  $ \sqrt{\frac{\varpi_k D_k c_k^3 R_k^2}{\eta}}-c_k R_k F_k^l \geq 0 $, i.e., $ \eta \leq \frac{\varpi_k D_k c_k}{{F_k^l} ^{2}} $. Since $ \eta \ne 0 $ and $\sum_{k=1}^{K} F_{k}^{e} $ monotonically decreases with respect to $\eta$, the optimal $\eta^{*}$ can be found in the range of $\left(\eta_l, \eta_u\right]=\left(0, \min _k\left(\frac{\varpi_k D_k c_k}{F_k^{l 2}}\right)\right]$ to ensure (\ref{Eq19}), using the classical bisection search method with the termination coefficient of $ \epsilon_{1} $. Algorithm 1 summarizes the procedure of solving problem (\ref{OptB}), whose computational complexity is mainly determined by calculating $F^{e(l_{1}+1)}$ using (\ref{Eq21}) and calculating $\eta$ using bisection search method. The computational complexity for calculating $ F^{e(l_{1}+1)} $ is $\mathcal{O}\left( \log _2\left(\frac{\eta_u-\eta_l}{\epsilon_{1}}\right) K\right)$. Hence the total complexity of Algorithm 1 is on the order of $\mathcal{O}\left(l_1^{\max } \log _2\left(\frac{\eta_u-\eta_l}{\epsilon_{1}}\right) K\right)$, where $ l_1^{\max } $ is the maximum number of iterations.
		
		\renewcommand{\algorithmicrequire}{\textbf{Input:}}
		\begin{algorithm}[t]
			\small
			\captionsetup{font={small}}
			\caption{Alternating Optimization of $\mathbf{d}$ and $\mathbf{F}^{e}$, Given $\mathbf{U}$ and $\boldsymbol{\theta}$ }\label{alg1}
			\begin{algorithmic}[0] 
				\Require 
				$ \mathbf{h}_{k} $, $ B $, $ p $, $ \sigma^{2} $, $ D_{k} $, $ c_{k} $, $ K $, $ \epsilon_{1} $, $l_{1}^{max}$, $ \mathbf{U} $, and $ \boldsymbol{\theta} $
				\Ensure
				Optimal $ \boldsymbol{d} $ and $ \mathbf{F}^{e} $ 
				\State \textbf{1. Initialization}\\
				Initialize $ l_{1}=0 $, calculate $ \left\lbrace R_{k} \right\rbrace_{k=1}^{K}  $ using (\ref{Eq8}) 
				\State \textbf{2. Alternating optimization of $ \boldsymbol{d} $ and $ \mathbf{F}^{e} $ }
				\Repeat:
				\State Update $ \eta $ using bisection search method
				\State Update $ \mathbf{F}^{e(l_{1}+1)} $ using (\ref{Eq21}) 
				\State Update $ \boldsymbol{d}^{(l_{1}+1)} $ using (\ref{Eq13})
				\State $ \epsilon_1^{\left(l_1+1\right)}=\frac{\left|\operatorname{obj}\left(\boldsymbol{d}^{\left(l_1+1\right)}, \mathbf{F}^{e\left(l_1+1\right)}\right)-\operatorname{obj}\left(\boldsymbol{d}^{\left(l_1\right)}, \mathbf{F}^{e\left(l_1\right)}\right)\right|}{\operatorname{obj}\left(\boldsymbol{d}^{\left(l_1+1\right)}, \mathbf{F}^{e\left(l_1+1\right)}\right)} $
				\State $ l_{1} \gets l_{1}+1$
				\Until{ $\epsilon_1^{\left(l_1+1\right)} \leq \epsilon_{1}$ $\|$  $l_{1} \geq l_{1}^{max} $ }
				\State \textbf{3. Return optimal $ \boldsymbol{d}^{*} $ and $ \mathbf{F}^{e*} $ } 
			\end{algorithmic}
		\end{algorithm}
		
		\subsection{Joint Optimization of BS's Receiving Vector and IRS BPS}			
		With a fixed offloading data volume $ \boldsymbol{d} $ and edge computing resource allocation $ \mathbf{F}^{e} $, problem (\ref{OptA}) can be reduced to
		\begin{subequations}\label{OptD}
			\begin{align}	
				& \min _{\mathbf{U}, \boldsymbol{\theta}} \sum_{k=1}^K \varpi_k T_k\left(\mathbf{u}_k, \boldsymbol{\theta}\right) \label{OptD1} \\
				&\text { s.t. } \text{(\ref{OptA2})}, \text{(\ref{OptA3})}, \text{(\ref{OptA4})}, \text{(\ref{OptA5})}
			\end{align}
		\end{subequations}
		\emph{Remark 2: } The challenges of solving problem (\ref{OptD}) include two aspects: 1) the segmented objective function form caused by the operation of maximization; 2) the objective function is the summation of fractional functions with respect to the BS's receiving vector $\mathbf{U}$ and BPS $ \boldsymbol{\theta} $. In order to tackle the first issue and solving the non-convex sum-of-ratios optimization problem caused by the second issue, we transform problem (\ref{OptD}) as follows.
		
		As explained in section \ref{IIIA}, when $ T_{k} $ takes its minimum value, we have $ T_{k}=T_{k}^{l}=T_{k}^{e} $. Hence we replace $ T_{k} $ with $ T_{k}^{e} $ and remove the constant terms, problem (\ref{OptD}) can be reformulated as 
		\begin{subequations}\label{OptE}
			\begin{align}
				& \min _{\mathbf{U}, \boldsymbol{\theta}} \sum_{k=1}^K \frac{\varpi_k d_k}{R_k\left(\mathbf{u}_k, \boldsymbol{\theta}\right)} \label{OptE1} \\
				& \text { s.t. (\ref{OptA2})-(\ref{OptA5})}.
			\end{align}
		\end{subequations}
		It can be converted into the following equivalent form by introducing auxiliary variable $ \boldsymbol{\xi} $
		\begin{subequations}\label{OptF}
			\begin{align}
				\min _{\mathbf{U}, \boldsymbol{\theta}, \boldsymbol{\xi}} & \sum_{k=1}^K \xi_k \\
				\text { s.t. } & \frac{\varpi_k d_k}{R_k\left(\mathbf{u}_k, \boldsymbol{\theta}\right)} \leq \xi_k,  \forall k, \\
				& \text{(\ref{OptA2})-(\ref{OptA5})}.
			\end{align}
		\end{subequations}
		The Lagrangian function related to problem (\ref{OptF}) is formulated as
		\begin{equation}
			\mathcal{L}(\mathbf{U}, \boldsymbol{\theta}, \boldsymbol{\xi}, \boldsymbol{\chi})\!=\!\sum_{k=1}^K \xi_k\!+\!\sum_{k=1}^K \chi_k\left[\varpi_k d_k\!-\!\xi_k R_k\left(\mathbf{u}_k, \boldsymbol{\theta}\right)\right],
		\end{equation}
		where $ \left\lbrace \chi_{k} \right\rbrace \geq 0  $ denotes the Lagrange multiplier. Corresponding, the augemented Lagrangian form of problem (\ref{OptF}) is given by 
		\begin{subequations}\label{OptG}
			\begin{align}
				& \min _{\mathbf{U}, \boldsymbol{\theta}, \boldsymbol{\xi},\boldsymbol{\chi} }  \mathcal{L}(\mathbf{U}, \boldsymbol{\theta}, \boldsymbol{\xi}, \boldsymbol{\chi}) \\
				& \text { s.t.  (\ref{OptA2})-(\ref{OptA5})} 
			\end{align}
		\end{subequations}
		Upon denoting the solution of problem (\ref{OptF}) by $ \left( \mathbf{U}^{*}, \boldsymbol{\theta}^{*}, \boldsymbol{\xi}^{*} \right)  $, there exist $ \boldsymbol{\chi}^{*} $ satisfying the following KKT conditions for $\forall k \in \mathcal{K}$
		\begin{align}
			& \frac{\partial \mathcal{L}}{\partial \theta_k}=-\chi_k^* \xi_k^* \nabla R_k\left(\mathbf{u}_k^*, \boldsymbol{\theta}^*\right)=0,  \label{Eq26}\\
			& \frac{\partial \mathcal{L}}{\partial \mathbf{u}_k}=-\chi_k^* \xi_k^* \nabla R_k\left(\mathbf{u}_k^*, \boldsymbol{\theta}^*\right)=0,  \label{Eq27}\\
			& \frac{\partial \mathcal{L}}{\partial \xi_k}=1-\chi_k^* R_k\left(\mathbf{u}_k^*, \boldsymbol{\theta}^*\right)=0,  \label{Eq28}\\
			& \chi_k^*\left[\varpi_k d_k-\xi_k^* R_k\left(\mathbf{u}_k^*, \boldsymbol{\theta}^*\right)\right]=0,  \label{Eq29}\\
			& \chi_k^* \geq 0, \label{Eq30}\\
			& \varpi_k d_k-\xi_k^* R_k\left(\mathbf{u}_k^*, \boldsymbol{\theta}^*\right) \leq 0,  \label{Eq31}\\
			& 0 \leq \theta_k^* \leq 2 \pi. \label{Eq32}
		\end{align}
		According to (\ref{Eq28}), (\ref{Eq29}), and $ R_k\left(\mathbf{u}_k^*, \boldsymbol{\theta}^*\right) \textgreater 0 $, we have
		\begin{equation}
			\chi_k^*=\frac{1}{R_k\left(\mathbf{u}_k^*, \boldsymbol{\theta}^*\right)}, \forall k ,
		\end{equation}
		and 
		\begin{equation}
			\xi_k^*=\frac{\varpi_k d_k}{R_k\left(\mathbf{u}_k^*, \boldsymbol{\theta}^*\right)}, \forall k.
		\end{equation}
		
		Next, problem (\ref{OptG}) will be solved in two steps: 1) obtain $ \mathbf{U}^{*} $ and $ \boldsymbol{\theta}^{*} $ by solving problem (\ref{OptG}) with fixed $ \boldsymbol{\xi} $ and $ \boldsymbol{\chi} $; 2) update $ \boldsymbol{\xi} $ and $ \boldsymbol{\chi} $ using a Newton-like method. $\left\lbrace \mathbf{U}^{*}, \boldsymbol{\theta}^{*} \right\rbrace $ and $\left\lbrace \boldsymbol{\xi}, \boldsymbol{\chi} \right\rbrace $ are alternately updated until objective (\ref{OptE1}) converges. The detailed procedures can refer to Algorithm \ref{alg2}. For convenience, we define
		\begin{align}
			& \Omega_k\left(\chi_k\right)=\chi_k R_k\left(\mathbf{u}_k^*, \boldsymbol{\theta}^*\right)-1, \forall k, \\
			& \psi_k\left(\xi_k\right)=\xi_k R_k\left(\mathbf{u}_k^*, \boldsymbol{\theta}^*\right)-\varpi_k d_k, \forall k.
		\end{align}
		\begin{algorithm}[ht]
			\small
			\captionsetup{font={small}}
			\caption{ Alternating Optimization $ \mathbf{U} $ and $ \boldsymbol{\theta} $, Given $ \boldsymbol{d} $ and $ \mathbf{F}^{e} $ }
			\label{alg2}
			\begin{algorithmic}[0]
				\Require
				$\boldsymbol{d}$, $ \mathbf{F}^{e} $, $\mathbf{h}_{k,p}^{d}$, $\mathbf{G}_{p}$, $\mathbf{h}_{k,p}^{r}$, $ \epsilon_{2}$ and $l_{2}^{\rm{max}}$, $\forall k \in \mathcal{K}$, $\forall p \in \mathcal{P}$ 
				\Ensure 
				Optimal $\mathbf{U}$ and $ \boldsymbol{\theta} $
				\State \textbf{1. Initialization}
				\State Initialize $ l_{2}=0 $, $ \tau \in \left(0, 1\right)  $, $ \Gamma \in \left(0, 1 \right) $, $ \boldsymbol{\theta} $  satisfying (\ref{OptA4}) and $ \mathbf{u}_{k,p} $ satisfying (\ref{OptA5}), $ \forall k \in \mathcal{K} $, $ \forall p \in \mathcal{P} $
				\State \textbf{2. Joint optimization $\mathbf{U}$, $\boldsymbol{\theta} $, $\boldsymbol{\chi}$, and $\boldsymbol{\xi}$ } 
				\While{$ \chi_{k}^{\left( l_{2} \right) }R_{k}\left( \mathbf{u}_{k}^{(l_{2})}, \boldsymbol{\theta}^{(l_{2})} \right)-1 > \epsilon_{2}  $    \& \&    $ \xi_{k}R_{k}\left( \mathbf{u}_{k}^{(l_{2})}, \boldsymbol{\theta}^{(l_{2})} \right)-\varpi_kd_{k} > \epsilon_{2}  $ \&\& $l_{2} \le l_{2}^{\rm{max}}$ }
				\State Update $ \mathbf{u}_{k,p}^{(l_{2}+1)} $ and $ \boldsymbol{\theta}^{(l_{2}+1)} $ using Algorithm \ref{alg3}, $ \forall k \in \mathcal{K} $, $ \forall p \in \mathcal{P} $
				\State  Update $ \boldsymbol{\chi}^{(l_{2}+1)} $ and $ \boldsymbol{\xi}^{(l_{2}+1)} $ as follows
				\State
				\begin{equation}      \chi_k^{\left(l_2+1\right)}=\chi_k^{\left(l_2\right)}-\frac{\Gamma^{i^{\left(l_2+1\right)}} \Omega_k\left(\chi_k^{\left(l_2\right)}\right)}{R_k\left(\mathbf{u}_k^{\left(l_2+1\right)}, \boldsymbol{\theta}^{\left(l_2+1\right)}\right)}
				\end{equation}
				\State  
				\begin{equation}
					\xi_{k}^{\left(l_2+1\right)}=\xi_{k}^{\left(l_2\right)}-\frac{\Gamma^{i^{\left(l_2+1\right)}} \psi_k\left(\xi_k^{\left(l_2\right)}\right)}{R_k\left(\mathbf{u}_k^{\left(l_2+1\right)}, \boldsymbol{\theta}^{\left(t_2+1\right)}\right)}
				\end{equation}
				\State where $ i^{(l_{2}+1)} $ is the smallest positive integer that satisfying
				\State 
				\begin{equation}
					\begin{aligned}
						&\sum_{k=1}^K\left|\Omega_k\left(\chi_k^{\left(l_2\right)}-\frac{\Gamma^i \Omega_k\left(\chi_k^{\left(l_2\right)}\right)}{R_k\left(\mathbf{u}_k^{\left(l_2+1\right)}, \boldsymbol{\theta}^{\left(l_2+1\right)}\right)}\right)\right|^2  \\
						& \quad \quad +\sum_{k=1}^K\left|\psi_k\left(\xi^{\left(l_2\right)}-\frac{\Gamma^i \psi_k\left(\xi^{\left(l_2\right)}\right)}{R_k\left(\mathbf{u}_k^{\left(l_2+1\right)}, \boldsymbol{\theta}^{\left(l_2+1\right)}\right)}\right)\right|^2 \\
						& \quad  \leq  \left(1-\tau \Gamma^i\right)^2 \sum_{k=1}^K\left[\left|\Omega_k\left(\chi_k^{\left(l_2\right)}\right)\right|^2+\left|\psi_k\left(\xi^{\left(l_2\right)}\right)\right|^2\right] 
					\end{aligned}
				\end{equation}
				\State $ l_{2}=l_{2}+1 $
				\EndWhile
				\State \textbf{3. Return $ \mathbf{u}_{k,p}^{*} $ and $ \boldsymbol{\theta}^{*} $, $ \forall k \in \mathcal{K} $, $ \forall p \in \mathcal{P} $}
			\end{algorithmic}
		\end{algorithm}
		
		Now, we focus our attention on the first step, i.e., optimizing $ \mathbf{U} $ and $ \boldsymbol{\theta} $. Given a set solution of $ \left\lbrace \boldsymbol{\chi},  \boldsymbol{\xi}, \boldsymbol{d} \right\rbrace $, problem (\ref{OptG}) can be reduced to the following weighted sum-rate maximization problem
		\begin{subequations}\label{OptH}
			\begin{align}
				&\underset{\mathbf{U}, \boldsymbol{\theta}}{\max } \sum_{k=1}^K  \sum_{p=1}^{P} \chi_k \xi_k R_{k,p}\left(\mathbf{u}_{k,p}, \boldsymbol{\theta}\right) \label{OptH1} \\
				& \text { s.t. (\ref{OptA2})-(\ref{OptA5})}.
			\end{align}
		\end{subequations}		
		Leveraging the equivalence between sum-rate maximization and MSE minimization \cite{ref40}, problem (\ref{OptH}) can be transformed into a modified MSE minimization problem. The latter is convex regarding each optimization variable while fixing others. Specifically, by introducing auxiliary variables $ \Upsilon_{k, p} \in \mathbb{C}, \forall k \in \mathcal{K}, \forall p \in \mathcal{P} $, the modified MSE function for the $k$-th WD on the $p$-th subcarrier can be formulated as 
		\begin{equation}
			\begin{aligned}
				\operatorname{MSE}_{k, p}= & \mathbb{E}\left\{\left(\Upsilon_{k, p}^* y_{k, p}-s_{k, p}\right)\left(\Upsilon_{k, p}^* y_{k, p}-s_{k, p}\right)^*\right\} \\
				= & \sum_{j=1}^K\left|\Upsilon_{k, p}^*\sqrt{\widetilde{p}}\mathbf{u}_{k, p}^{H} \left( \mathbf{h}_{j, p}^{\mathrm{d}}+\mathbf{G}_p\boldsymbol{\Phi}_p \mathbf{h}_{j, p}^{\mathrm{r}}  \right)  \right|^2 \\
				& -2 \Re\left\{\Upsilon_{k, p}^*\sqrt{\widetilde{p}}\mathbf{u}_{k, p}^{H}\left( \mathbf{h}_{k, p}^{\mathrm{d}}+\mathbf{G}_p\boldsymbol{\Phi}_p \mathbf{h}_{k, p}^{\mathrm{r}}  \right)  \right\} \\
				& +\left|\Upsilon_{k, p}\right|^2  \sigma^2+1, \quad \forall k, p.
			\end{aligned}
		\end{equation}  
		Next, problem (\ref{OptH}) can be converted into the following equivalent form by introducing weighting parameters $ \rho_{k,p} \in \mathbb{R}^{+}, \forall k \in \mathcal{K}, \forall p \in \mathcal{P} $
		\begin{subequations}\label{OptI}
			\begin{align}
				& \underset{\mathbf{U}, \boldsymbol{\theta}, \boldsymbol{\Upsilon}, \boldsymbol{\rho}}{\max } \sum_{k=1}^K \sum_{p=1}^P \chi_k \xi_k \left(\log _2 \rho_{k, p}-\rho_{k, p} \mathrm{MSE}_{k, p}+1\right) \label{OptI1}\\
				& \text { s.t. (\ref{OptA2})-(\ref{OptA5})},
			\end{align}
		\end{subequations}
		where $ \boldsymbol{\Upsilon} $ and $ \boldsymbol{\rho} $ represent the sets of variables $ \Upsilon_{k, p} $ and $ \rho_{k,p} $, respectively. Compared with problem (\ref{OptH}), the newly formulated problem (\ref{OptI}) is more tractable, because the complex SINR term is removed from the $\log(\cdot)$ function. In the following subsection, we will decouple problem (\ref{OptI}) into four subproblems and provide a solution for each subproblem separately. 
		
		\emph{1) Optimization of Weighting Parameter $ \boldsymbol{\rho} $:} Given the BS's receiving vector $\mathbf{U}$, BPS $\boldsymbol{\theta}$ and auxiliary variable $\boldsymbol{\Upsilon}$, the subproblem with respect to weighting parameters $ \rho_{k,p} $ is formulated as 
		\begin{equation}\label{OptJ}
			\underset{\rho_{k,p}}{\max } \quad \chi_k \xi_k \left( \log _2 \rho_{k, p}-\rho_{k, p} \mathrm{MSE}_{k, p}\right), 
		\end{equation}
		which is a convex problem without constraints, and its solution is provided by setting the first-order partial derivative of objective function (\ref{OptJ}) with respect to $ \rho_{k,p} $ to zero, which is expressed as
		\begin{equation}\label{Eq42}
			\rho_{k, p}^{\star}=\mathrm{MSE}_{k, p}^{-1}=1+\gamma_{k, p},  \forall k, p
		\end{equation}
		
		\begin{figure*}[t]
			\begin{equation}
				\begin{aligned}
					& g_{1}=\sum_{k=1}^K \sum_{p=1}^P \chi_k \xi_k \rho_{k, p}  \left(   \sum_{j=1}^K\left|\Upsilon_{k, p}^*\sqrt{\widetilde{p}}\mathbf{u}_{k, p}^{H} \mathbf{h}_{j,p}  \right|^2  -2 \Re\left\{\Upsilon_{k, p}^*\sqrt{\widetilde{p}}\mathbf{u}_{k, p}^{H}\mathbf{h}_{k,p}  \right\} \right)    \\
					&\quad = \sum_{k=1}^K \sum_{p=1}^P  \left[  \mathbf{u}_{k, p}^{H} \left( \sum_{j=1}^K \chi_k \xi_k \rho_{k, p} \widetilde{p} \left|\Upsilon_{k, p}\right|^{2} \mathbf{h}_{j,p} \mathbf{h}_{j,p}^{H} \right) \mathbf{u}_{k, p}\right]   - \sum_{k=1}^K \sum_{p=1}^P 2\chi_k \xi_k \rho_{k, p}\Re\left\lbrace \Upsilon_{k, p}^*\sqrt{\widetilde{p}}\mathbf{u}_{k, p}^{H}\mathbf{h}_{k,p} \right\rbrace.
				\end{aligned}
			\end{equation}
			{\noindent}
			\rule[-10pt]{18.07cm}{0.1em}
		\end{figure*}
		
		\emph{2) Optimization of Auxiliary Variable $ \boldsymbol{\Upsilon} $:} Fixing BS's receiving vector $\mathbf{U}$, BPS $\boldsymbol{\theta}$ and weighting parameters $ \boldsymbol{\rho} $, the subproblem with respect to the auxiliary variable $ \Upsilon_{k,p} $ can be written as
		\begin{subequations}\label{OptK}
			\begin{align}
				\underset{\Upsilon_{k,p}}{\min } \quad \chi_k \xi_k \rho_{k, p} \mathrm{MSE}_{k, p}, \forall k, p,
			\end{align}
		\end{subequations}
		similarly, we select $\Upsilon_{k, p}^{\star}$ which ensures the partial derivative with respect to $ \Upsilon_{k,p} $ to zero as its optimal solution, i.e.,
		\begin{equation}\label{Eq44}
			\Upsilon_{k, p}^{\star}=\dfrac{\sqrt{\widetilde{p}}\mathbf{u}_{k,p}^{H}\left( \mathbf{h}_{k,p}^{d} +\mathbf{G}_{p} \mathbf{\Phi}_{p} \mathbf{h}_{k,p}^{r}  \right) }{ \sum_{j=1}^{K}\left|\sqrt{\widetilde{p}}\mathbf{u}_{k,p}^{H}\left(\mathbf{h}_{j,p}^{d}  +\mathbf{G}_{p}\mathbf{\Phi}_{p} \mathbf{h}_{j,p}^{r}   \right) \right|^{2}+\sigma^{2} },  \forall k, p
		\end{equation}
		
		\emph{3) Optimization of BS's receiving Vector $ \mathbf{U} $:} Given the BPS $\boldsymbol{\theta}$, weighting parameters $ \boldsymbol{\rho} $, and auxiliary variable $ \boldsymbol{\Upsilon} $, the subproblem with respect to the BS's receiving vector $\mathbf{U}$ can be expressed as
		\begin{subequations}\label{OptL}
			\begin{align}
				& \underset{\mathbf{U}}{\min } \! \sum_{k=1}^K \sum_{p=1}^P \chi_k \xi_k \rho_{k, p}  [ \sum_{j=1}^K\left|\Upsilon_{k, p}^*\sqrt{\widetilde{p}}\mathbf{u}_{k, p}^{H} \left( \mathbf{h}_{j, p}^{\mathrm{d}}\!+\!\mathbf{G}_p\boldsymbol{\Phi}_p \mathbf{h}_{j, p}^{\mathrm{r}}  \right) \! \right|^2 \notag \\
				& \quad \quad  -2 \Re\left\{\Upsilon_{k, p}^*\sqrt{\widetilde{p}}\mathbf{u}_{k, p}^{H}\left( \mathbf{h}_{k, p}^{\mathrm{d}}+\mathbf{G}_p\boldsymbol{\Phi}_p \mathbf{h}_{k, p}^{\mathrm{r}}  \right)  \right\} ] \label{OptL1} \\
				& \text { s.t.}  \left\| \mathbf{u}_{k,p} \right\|^{2} \leq 1. \label{OptL2}
			\end{align}
		\end{subequations}
		
		To simplify the expression, we define equivalent channel for the $k$-th WD on the $p$-th subcarrier as $ \mathbf{h}_{k,p}= \mathbf{h}_{k, p}^{\mathrm{d}}+\mathbf{G}_p\boldsymbol{\Phi}_p \mathbf{h}_{k, p}^{\mathrm{r}}, \forall k, p $. The objective  (\ref{OptL1}) can be simplified as $g_{1}$, which is presented at the top of this page.
		
		To further simplify the expression of $ g_{1} $, we define
		\begin{align}
			& \mathbf{a}_{k,p}=\sum_{j=1}^K \chi_k \xi_k \rho_{k, p} \widetilde{p} \left|\Upsilon_{k, p}\right|^{2} \mathbf{h}_{j,p} \mathbf{h}_{j,p}^{H}, \label{Eqa}\\
			& \mathbf{A}_{p}\!=\!\operatorname{diag}\left[ \mathbf{a}_{1,p},...,\mathbf{a}_{K,p} \right], 
			\mathbf{A}\!=\!\operatorname{diag}\left(\mathbf{A}_1, \cdots, \mathbf{A}_P\right), \label{Eqb}\\
			&\mathbf{v}_{k,p}=\chi_k \xi_k \rho_{k, p} \Upsilon_{k, p}^*\sqrt{\widetilde{p}} \mathbf{h}_{k,p},  \label{Eqc}\\
			&\mathbf{V}=\left[\mathbf{v}_{1,1}^{T},\mathbf{v}_{2,1}^{T},..., \mathbf{v}_{K,1}^{T},\mathbf{v}_{1,2}^{T},\mathbf{v}_{2,2}^{T},...,\mathbf{v}_{K, P}^{T} \right]^{T}, \label{Eqd}
		\end{align}
		By substituting (\ref{Eqa})-(\ref{Eqd}) into $ g_{1} $, $ g_{1} $ can be rewritten as
		\begin{equation}
			\begin{aligned}
				g_{2}\left( \mathbf{U} \right)= \mathbf{U}^{H}\mathbf{A}\mathbf{U}-\Re\left\lbrace 2 \mathbf{U}^{H}\mathbf{V} \right\rbrace.
			\end{aligned}
		\end{equation}
		So far, problem (\ref{OptL}) is reduced to
		\begin{subequations}\label{OptM}
			\begin{align}
				&\underset{\mathbf{U}}{\min } \quad \mathbf{U}^{H}\mathbf{A}\mathbf{U}-\Re\left\lbrace 2 \mathbf{U}^{H}\mathbf{V} \right\rbrace \label{OptM1} \\
				&\text { s.t.}  \mathbf{U}^{H}\mathbf{D}_{k,p}\mathbf{U} \leq 1, \forall k, p, \label{OptM2}
			\end{align}
		\end{subequations}
		where $ \mathbf{D}_{k,p}=\mathbf{O}_{k,p} \otimes \mathbf{I}_{M} $, $ \mathbf{O}_{k,p} $ is a $ K \times P $ matrix whose $ \left( k, p \right) $-th element is one, and other elements are zeros. Since matrices $ \mathbf{A} $ and $ \mathbf{D}_{k, p} $ are all positive semidefinite, the simplified subproblem ($ \ref{OptM} $) is a standard convex optimization problem, which can be optimally solved using the existing convex optimization toolbox \cite{ref41}.
		
		\emph{4) Optimization of BPS Vector $ \boldsymbol{\theta} $:} When the BS's receiving vector $\mathbf{U}$, weighting parameters $ \boldsymbol{\rho} $, and auxiliary variable $ \boldsymbol{\Upsilon} $ are all fixed, the subproblem regarding BPS $\boldsymbol{\theta}$ can be written as
		\begin{subequations}\label{OptN}
			\begin{align}
				& \underset{\boldsymbol{\theta}}{\min } \! \sum_{k=1}^K \sum_{p=1}^P \chi_k \xi_k \rho_{k, p}  \Bigg( \sum_{j=1}^K\left|\Upsilon_{k, p}^*\sqrt{\widetilde{p}}\mathbf{u}_{k, p}^{H} \left( \mathbf{h}_{j, p}^{\mathrm{d}}\!+\!\mathbf{G}_p\boldsymbol{\Phi}_p \mathbf{h}_{j, p}^{\mathrm{r}}  \right)  \right|^2 \notag\\
				&\qquad -2 \Re\left\{\Upsilon_{k, p}^*\sqrt{\widetilde{p}}\mathbf{u}_{k, p}^{H}\left( \mathbf{h}_{k, p}^{\mathrm{d}}+\mathbf{G}_p\boldsymbol{\Phi}_p \mathbf{h}_{k, p}^{\mathrm{r}}  \right)  \right\} \Bigg)  \\
				& \text { s.t. (\ref{OptA2})-(\ref{OptA4})}.
			\end{align}
		\end{subequations}
		Upon defining $ \boldsymbol{\phi}_{p} \triangleq\left[\phi_{p, 1}, \ldots, \phi_{p, N}\right]^T $, $ \bar{h}_{k, j, p}^{\mathrm{d}} \triangleq \mathbf{u}_{k, p}^{H}\mathbf{h}_{j, p}^{\mathrm{d}}  $ and $\mathbf{g}_{k, j, p} \triangleq   \mathbf{u}_{k,p}^{H} \mathbf{G}_{p} \operatorname{diag}\left( \mathbf{h}_{j,p}^{r}\right) , \forall k, j \in \mathcal{K}, \forall p \in \mathcal{P}$, problem (\ref{OptN}) can be reformulated as (\ref{OptO}), which is presented at the top of this page.
		\begin{figure*}[t]
			\begin{subequations}\label{OptO}
				\begin{align}
					& \underset{\boldsymbol{\theta}}{\min} \sum_{p=1}^P \sum_{k=1}^K \chi_{k} \xi_k \rho_{k, p}\!\!\left(\sum_{j=1}^K\left|\Upsilon_{k, p}^* \sqrt{\widetilde{p}}\left(\bar{h}_{k, j, p}^{\mathrm{d}}+\mathbf{g}_{k, j, p} \boldsymbol{\phi}_p \right)\right|^2-2 \Re\left\{\Upsilon_{k, p}^*\sqrt{\widetilde{p}}\left(\bar{h}_{k, k, p}^{\mathrm{d}}+\mathbf{g}_{k, k, p} \boldsymbol{\phi}_p\right)\right\}\right. \Bigg) \label{OptO1}\\
					&= \underset{\boldsymbol{\Theta}}{\min}  \sum_{p=1}^P\left(\boldsymbol{\phi}_p^H \boldsymbol{\Lambda}_p \boldsymbol{\phi}_p-2 \Re\left\{ \boldsymbol{\phi}_p^{H}\boldsymbol{\nu}_p\right\}+\zeta_{p}\right) \label{OptO2}\\
					&\text { s.t. } \rm{(\ref{OptA2})}-\rm{(\ref{OptA4})},
				\end{align}
			\end{subequations}
			{\noindent}
			\rule[-10pt]{18.07cm}{0.1em}	
		\end{figure*}	
		Matrix $ \boldsymbol{\Lambda}_{p} $, vector $ \boldsymbol{\nu}_{p} $ and scalar $ \zeta_{p} $, $ \forall p \in \mathcal{P} $, are respectively defined as (\ref{Eq55}), (\ref{Eq56}), and (\ref{Eq57}), which are presented at the bottom of this page.
		\begin{figure*}[b]
			{\noindent}
			\rule[-10pt]{18.07cm}{0.1em}
			\begin{align}
				&\boldsymbol{\Lambda}_p \!\triangleq \! \sum_{k=1}^K \chi_{k} \xi_k \rho_{k, p}\left|\Upsilon_{k, p}\right|^2 \widetilde{p} \sum_{j=1}^K \mathbf{g}_{k, j, p}^{H} \mathbf{g}_{k, j, p} , \forall p, \label{Eq55}\\
				&\boldsymbol{\nu}_p \! \triangleq \! \sum_{k=1}^K \chi_{k} \xi_k \rho_{k, p} \! \left(\!\Upsilon_{k, p}\sqrt{\widetilde{p}} \mathbf{g}_{k, k, p}^{H}\!-\!\left|\Upsilon_{k, p}\right|^2 \!\widetilde{p}\!\sum_{j=1}^K \mathbf{g}_{k, j, p}^{H} \bar{h}_{k, j, p}^{d} \! \right), \forall p \label{Eq56}\\
				&\zeta_{p} \! \triangleq \! \sum_{k=1}^{K} \chi_{k} \xi_k \rho_{k, p} \! \left( \! \left|\Upsilon_{k, p}\right|^2 \!\widetilde{p}\! \sum_{j=1}^{K} \left|  \bar{h}_{k, j, p}^{d} \right| ^{2} \!\!\!-\!2 \Re \left\lbrace \Upsilon_{k, p}^{*} \sqrt{\widetilde{p}} \bar{h}_{k, k, p}^{d} \right\rbrace \! \right), \forall p  \label{Eq57}
			\end{align}
		\end{figure*}	
		
		It is challenging to solve problem (\ref{OptO}) directly because of the BPS $\boldsymbol{\theta}$ is embedded into $P$ complicated functions. To handle this issue, one feasible scheme is to decouple the original problem to $N$ subproblems, only one element of the BPS $\boldsymbol{\theta}$ is optimized each time.
		
		To proceed, the objective function (\ref{OptO2}) is divided element-by-element as
		\setlength{\abovedisplayskip}{3pt}
		\begin{align}
			&\sum_{p=1}^P\left(\boldsymbol{\phi}_p^H \boldsymbol{\Lambda}_p \boldsymbol{\phi}_p-2 \Re\left\{\boldsymbol{\phi}_p^H \boldsymbol{\nu}_p\right\}\right) \label{Eq61}\\
			&=\sum_{p=1}^P \sum_{n=1}^N\left(\sum_{n^{\prime}=1}^N \boldsymbol{\Lambda}_p(n, n^{\prime}) \phi_{p, n}^* \phi_{p, n^{\prime}}-2 \Re\left\{\phi_{p, n}^* \boldsymbol{\nu}_p(n)\right\}\right)\!.\notag
		\end{align}	
		Fixing the other elements, the objective function regarding the $n$-th element can be expressed as
		\begin{equation}
			\begin{aligned}
				&\sum_{p=1}^P( \sum_{n^{\prime} \neq n}\left(\boldsymbol{\Lambda}_p(n, n^{\prime}) \phi_{p, n}^* \phi_{p, n^{\prime}}+\boldsymbol{\Lambda}_p(n^{\prime}, n) \phi_{p, n^{\prime}}^* \phi_{p, n}\right) \\
				& \left.+\boldsymbol{\Lambda}_p(n, n)\left|\phi_{p, n}\right|^2-2 \Re\left\{\phi_{p, n}^* \boldsymbol{\nu}_p(n)\right\}\right) \\
				\stackrel{(\mathrm{a})}{=} & \sum_{p=1}^P\left(\sum_{n^{\prime} \neq n}\left(\boldsymbol{\Lambda}_p(n, n^{\prime}) \phi_{p, n}^* \phi_{p, n^{\prime}}+\boldsymbol{\Lambda}_p^*(n, n^{\prime}) \phi_{p, n} \phi_{p, n^{\prime}}^*\right)\right. \\
				& \left.+\boldsymbol{\Lambda}_p(n, n)\left|\phi_{p, n}\right|^2-2 \Re\left\{\phi_{p, n}^* \boldsymbol{\nu}_p(n)\right\}\right) \\
				= & \sum_{p=1}^P\left(2 \Re\left\{\left(\sum_{n^{\prime} \neq n} \boldsymbol{\Lambda}_p(n, n^{\prime}) \phi_{p, p^{\prime}}-\boldsymbol{\nu}_p(n)\right) \phi_{p, n}^*\right\}\right. \\
				& \left.+\boldsymbol{\Lambda}_i(n, n)\left|\phi_{p, n}\right|^2\right),
			\end{aligned}
		\end{equation}
		in which (a) holds since $ \boldsymbol{\Lambda}_{p}=\boldsymbol{\Lambda}_{p}^{H}, \forall p \in \mathcal{P} $. The subproblem with respect to the $n$-th element $\theta_{n}$ is given by
		\begin{subequations}\label{OptP}
			\begin{align}
				&\min _{\theta_n} \sum_{p=1}^P\left(2 \Re\left\{\left(\sum_{n^{\prime} \neq n} \boldsymbol{\Lambda}_p(n, n^{\prime}) \phi_{p, n^{\prime}}-\boldsymbol{\nu}_p(n)\right) \phi_{p, n}^*\right\}\right. \notag \\
				&\quad \quad \left.+\boldsymbol{\Lambda}_p(n, n)\left|\theta_{p, n}\right|^2\right) \label{OptP1}\\
				&\text{ s.t. (\ref{OptA2})-(\ref{OptA4}). }
			\end{align}
		\end{subequations}
		For convenience, we further define $\omega_{p, n} \triangleq \sum_{n^{\prime} \neq n} \boldsymbol{\Lambda}_p(n, n^{\prime}) \phi_{p, n^{\prime}}-\boldsymbol{\nu}_p(n), \forall p \in \mathcal{P}, \forall n \in \mathcal{N}$. Upon substituting constraints (\ref{OptA2}) and (\ref{OptA3}) into objective (\ref{OptP1}), problem (\ref{OptP}) can be rewritten as
		\begin{subequations}\label{OptQ}
			\begin{align}
				\min _{\theta_n} & \quad g_{3} \left( \theta_n \right) \label{OptQ1}\\
				\text { s.t. } & \theta_n \in[-\pi, \pi], \forall n,  \label{OptQ2}
			\end{align}
		\end{subequations}
		where  
		\begin{align}\label{Eq62}
			g_{3}\left( \theta_{n} \right)=&\sum_{p=1}^P \Big(2\left|\omega_{p, n}\right| \mathrm{A}\left(\theta_n, f_p\right) \cos \left(\angle \omega_{p, n}-\mathrm{B}\left(\theta_n, f_p\right)\right)\notag\\
			&+\boldsymbol{\Lambda}_p(n, n) \mathrm{A}^2\left(\theta_n, f_p\right)\Big).
		\end{align}
		\setlength{\belowdisplayskip}{3pt}
		Problem (\ref{OptQ}) is still difficult to solve because the complicated objective function (\ref{OptQ1}). A large number of numerical simulations indicates that function (\ref{OptQ1}) behaves like a smooth curve with two peak-trough. Thereby, there is only one minimum point within the range $\left[-\pi, \pi \right] $. Inspired by this observation, we adopt the following three-phase search method to search its optimal solution.
		
		\noindent
		\hangafter=1
		\setlength{\hangindent}{4em}
		\emph{Step 1:} We first focus our attention on narrowing the search range. Initialize the start point within the range $\left[-\pi, \pi \right] $ and search step length $l \textgreater 0 $. If $g_{3}\left( \theta_{n}^{0}+l\right) \textless g_{3}\left(\theta_{n}^{0} \right) $, search forward; otherwise search reversely, until the value of $g_{3}$ rises. 
		
		\noindent
		\hangafter=1
		\setlength{\hangindent}{4em}
		\emph{Step 2:} Determine the minimum point $\widetilde{\theta}$ by continuously sectioning the search range, until a predetermined threshold is met.
		
		\noindent
		\hangafter=1
		\setlength{\hangindent}{4em}
		\emph{Step 3:} The minimum value of $g_{3}$ and the corresponding phase shift is determined by comparing the objective value at the points $\widetilde{\theta}_{n}$, $-\pi$ and $\pi$.
		
		In order to reduce hardware consumption and cost, we can use finite discrete phases to realized IRSs in realistic applications. Assuming that these discrete phases of BPS $\theta_{n}$ is controlled by $b$ bits, which is given by
		\begin{equation}\label{Eq63}
			\theta_n \in \mathcal{S} \triangleq\left\{\frac{2 \pi}{2^b} i-\pi \mid i=0,1, \ldots, 2^b-1\right\},  \forall n.
		\end{equation}
		Accordingly, the IRS optimization subproblem for discrete phases is expressed as
		\begin{subequations}\label{OptR}
			\begin{align}
				\min _{\theta_n} & \sum_{p=1}^{P}\left(2\left|\omega_{p, n}\right| \mathrm{A}\left(\theta_n, f_p\right) \cos \left(\angle \omega_{p, n}-\mathrm{B}\left(\theta_n, f_p\right)\right)\right. \notag\\
				& \left.+\boldsymbol{\Lambda}_p(n, n) \mathrm{A}^2\left(\theta_n, f_p\right)\right) \label{OptR1}\\
				\text { s.t. }& \theta_n \in \mathcal{S}, \forall n.  \label{OptR2}
			\end{align}
		\end{subequations}
		It can be seen that except for constraint (\ref{OptR2}), the discrete IRS design subproblem is almost identical to that of continuous phases case. To find the optimal $\boldsymbol{\theta}$, the one-dimensional exhaustive search can be performed over the feasible set $\mathcal{S}$ due to the application of low-resolution phase shifter. Specifically, we alternately optimize each element in the BPS $\boldsymbol{\theta}$, while fixing the others until convergence. To determine the optimal BPS $\theta_{n}^{*}$, we check all the feasible solution from set $\mathcal{S}$ and calculate the corresponding reflection coefficients for each subcarrier using the non-ideal phase shift model. Subsequently, the optimal $\theta_{n}^{*}$ is selected. 
		
		Up to now, we provide the detailed approaches for solving the above four subproblems with respect to $ \mathbf{U} $, $ \boldsymbol{\theta} $, $ \boldsymbol{\Upsilon} $ and $ \boldsymbol{\rho} $, the overall procedure for solving problem (\ref{OptI}) is summarized in Algorithm \ref{alg3}. Given appropriate initial values for $\mathbf{U}$ and $\boldsymbol{\theta}$, we alternately update the four variables until the objective converges.
		\begin{algorithm}[ht]
			\small
			\captionsetup{font={small}}
			\caption{Alternating Optimization $ \mathbf{U} $ and $ \boldsymbol{\theta} $, Given $ \boldsymbol{\chi} $ and $ \boldsymbol{\xi} $}
			\label{alg3}
			\begin{algorithmic}[0]
				\Require
				$B$, $\mathbf{h}_{k,p}^{d}$, $\mathbf{G}_{p}$, $\mathbf{h}_{k,p}^{r}$, $ \forall p \in \mathcal{P} $, $\forall k \in \mathcal{K} $
				\Ensure
				$\mathbf{U}$ and $\boldsymbol{\theta}$ 
				\State \textbf{1. Initialization} 
				\State $ \boldsymbol{\theta} $ and $ \mathbf{u}_{k,p} $, $ \forall p \in \mathcal{P} $, $\forall k \in \mathcal{K} $
				\State \textbf{2. Alternating optimization $ \mathbf{U} $, $\boldsymbol{\theta}$, $\boldsymbol{\rho}$, $\boldsymbol{\Upsilon}$ } 
				\Repeat:
				\State Update $ \rho_{k,p} $ using (\ref{Eq42}), $ \forall k \in \mathcal{K} $, $ \forall p \in \mathcal{P} $
				\State Update $ \Upsilon_{k,p} $ using (\ref{Eq44})$, \forall k \in \mathcal{K} $, $ \forall p \in \mathcal{P} $
				\State Update $ \mathbf{u}_{k,p} $ by solving problem (\ref{OptM}), $ \forall k \in \mathcal{K} $, $ \forall p \in \mathcal{P} $
				\Repeat
				\For{$ n=1 $ to $ N $}
				\State Continuous phase case: update $ \theta_{n} $ by three-phase 
				\State one-dimensional search method
				\State Low-resolution phases case: update $ \theta_{n} $ by exhaustive 
				\State search
				\EndFor
				\Until{BPS $ \boldsymbol{\theta} $ converges}
				\Until{objective (\ref{OptI1}) converges}
				\State \textbf{3. Return optimal $ \mathbf{u}_{k,p}^{*} $ and $ \boldsymbol{\theta}^{*} $, $ \forall k \in \mathcal{K} $, $ \forall p \in \mathcal{P} $}
			\end{algorithmic}
		\end{algorithm}
		
		\begin{algorithm*}[htp]
			\small
			\captionsetup{font={small}}
			\caption{ The Method for Solving Problem (\ref{OptA})}
			\label{alg4}
			\begin{algorithmic}[0]
				\Require
				$D_{k}$, $ c_{k} $, $ F_{\rm{total}}^{e} $, $ \epsilon $, and $ l_{3}^{\max} $
				\Ensure
				Optimal $ \boldsymbol{d} $, $ \mathbf{F}^{e} $, $\mathbf{u}_{k,p} $ and $ \boldsymbol{\theta} $, $ \forall k \in \mathcal{K},  \forall p \in \mathcal{P}  $ 
				\State \textbf{1. Initialization} \\
				Initialize $ l_{3}=0 $ and $ \epsilon_{3}^{\left( 0 \right) }=1 $ \\
				Initialize $ \boldsymbol{\theta}^{(0)} $ satisfying (\ref{OptA4}), $ \mathbf{u}_{k,p}^{(0)} $ satisfying (\ref{OptA5}), and $ \mathbf{F}^{e(0)} $ satisfying (\ref{OptA7}) and (\ref{OptA8}) 
				\State \textbf{2. Alternating optimization of $ \boldsymbol{d} $ and $ \mathbf{F}^{e} $, $ \mathbf{u}_{k,p}^{(l_{3})} $ and $ \boldsymbol{\theta}^{(l_{3})} $}
				\Repeat
				\State Updating $ \boldsymbol{d}^{(l_{3}+1)} $ and $ \mathbf{F}^{e(l_{3}+1)} $ using Algorithm \ref{alg1}
				\State Updating $ \mathbf{u}_{k,p}^{(l_{3}+1)} $ and $ \boldsymbol{\theta}^{(l_{3}+1)} $ using Algorithm \ref{alg2}
				\State $\epsilon_3^{\left(l_3\right)}=\frac{\operatorname{obj}\left(\mathbf{d}^{\left(l_3+1\right)}, \mathbf{F}^{e\left(l_3+1\right)}, \mathbf{U}^{\left(l_3+1\right)}, \boldsymbol{\theta}^{\left(l_3+1\right)}\right)-\operatorname{obj}\left(\mathbf{d}^{\left(l_3\right)}, \mathbf{F}^{e\left(l_3\right)}, \mathbf{U}^{\left(l_3\right)}, \boldsymbol{\theta}^{\left(l_3\right)}\right) \mid}{\operatorname{obj}\left(\mathbf{d}^{\left(l_3+1\right)}, \mathbf{F}^{e\left(l_3+1\right)}, \mathbf{U}^{\left(l_3+1\right)}, \boldsymbol{\theta}^{\left(l_3+1\right)}\right)}$
				\State $l_3=l_3+1$
				\Until{$ \epsilon_3^{\left(l_3\right)} < \epsilon $ $ \| \| $ $ l_{3} > l_{3}^{\max} $}
				\State 	\textbf{3. Output the optimal $ \boldsymbol{d}^{*} $, $ \mathbf{F}^{e*} $, $ \mathbf{u}_{k,p}^{*} $ and $ \boldsymbol{\theta}^{*} $, $ \forall k \in \mathcal{K} $, $ \forall p \in \mathcal{P} $}
			\end{algorithmic}
		\end{algorithm*}
		
		\subsection{Complexity Analysis of Algorithm 2}
		Since the update of $ \boldsymbol{\chi} $ and $ \boldsymbol{\xi} $ both have explicit mathematical expressions, the computational complexity of Algorithm \ref{alg2} is mainly dominated by jointly solving $ \mathbf{U} $ and $ \boldsymbol{\theta} $. Specifically, in each iteration, updating weighting parameter $ \boldsymbol{\rho} $, auxiliary variable $ \boldsymbol{\Upsilon} $, and BS's receiving vector $ \mathbf{U} $ require about $ \mathcal{O}\left( PK^{2}MN^{2} \right)  $, $ \mathcal{O}\left( PK\left( K+1 \right) M N^{2}  \right)  $, and $ \mathcal{O}\left(I_{1}\left( M^{2}P^{2}K^{2}+MPK\right) \right)  $ operations, respectively.
		Finally, updating BPS $ \boldsymbol{\theta} $ for continuous phases is $ \mathcal{O}\left( \left( 5NM+N^{3} \right)PK^{2}+I_{2}PN\left( I_{3}+I_{4}\right) \right)  $, where $ I_{2} $ denotes the numbers of iterations for calculating $ \boldsymbol{\theta} $, $ I_{3} $ and $ I_{4} $ denote the iteration for step 2 and step 3 of the three-phase search method, respectively. For the discrete phases, the order of complexity for updating $ \boldsymbol{\theta} $ is about $ \mathcal{O}\left( \left( 5NM+N^{3} \right)PK^{2}+I_{5}PN2^{b}  \right)  $, where $ I_{5} $ denotes the number of iterations for calculating BPS $ \boldsymbol{\theta} $. 
		
		\section{Numerical Results}
		\begin{figure}[htbp]
			\begin{minipage}[t]{0.42\textwidth}
				\centering
				\includegraphics[width=0.9\textwidth]{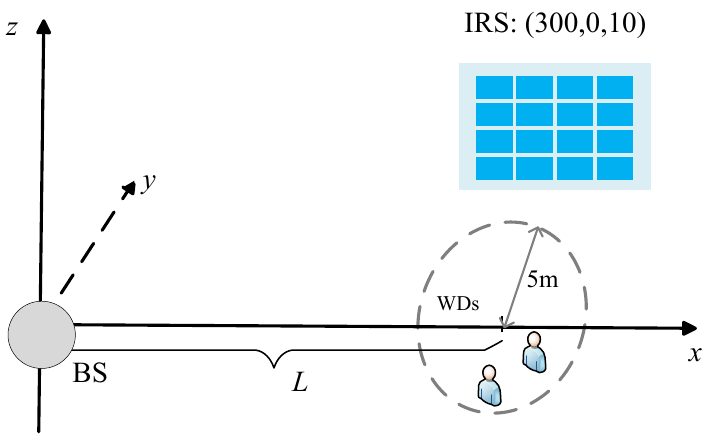}
				\caption{The simulation scenario setup.}
				\label{fig9}
			\end{minipage}
		\end{figure}
		In this section, we will evaluate the proposed algorithm by a large number of simulation results. As illustrated in Fig. \ref{fig9}, we consider a single-cell MEC system in which the BS's coverage radius is 300 meters \cite{ref24}. To improve the communication capacity, an IRS is deployed at the cell edge, which is high enough to construct a additional reflection link to assist the data offloading. Two WDs are randomly distributed in a circle centered at $\left(L, 0, 0 \right) $ m with radius $r=5$ m. The detailed location information for BS and IRS is presented in the “Location setting” block of Table \!\ref{tab2}.
		
		\begin{table}[t]\small
			\centering
			\captionsetup{font={small}}
			\caption{\quad System Parameters}
			\begin{tabular}{|l|r|}
				\hline
				Description & Parameter and Value \\
				\hline
				{Location setting} & \tabincell{r}{ BS: (0 m, 0 m, 0m) \\ IRS: (300m, 0m, 10 m)\\ $L=290 $ m, $ r=5 $ m} \\ 
				\hline
				{Communication setting}  & \tabincell{r}{ $K=2$, $ \widetilde{p}=1 $ mW, $ M=4 $\\$ N=20 $,  $ f_{c}=2.4 $ GHz \\ $ B=100 $ MHz, $ P=8 $, $ b=3 $ \\  $ \varpi_{k}=1/K $, $\sigma^{2}=3.98 \times 10^{-15}$ W \\ $ \Gamma=0.5 $, $ \tau=0.1 $ \\  $ \kappa_{\rm{BW}}=3.5 $, $ \kappa_{\rm{BI}}=2.2 $, $ \kappa_{\rm{IW}}=2.2 $  } \\
				\hline
				{Computing setting}  & \tabincell{r}{$ D_{k}=\left[250, 350 \right] \times 10^{3} $ bit\\ $ c_{k}=\left[700,800 \right]  $ cycle/bit \\ $ F_{k}^{l}=\left[ 4,6\right] \times 10^{8} $ cycle/s\\  $ F_{\rm{total}}^{e}=50\times 10^{11} $ cycle/s } \\
				\hline
			\end{tabular}
			\label{tab2}
		\end{table}
		
		We assume that the number of subcarriers is $P=8$. The carrier frequency and bandwidth setting follows \cite{ref35} and are given by $ f_{c}=2.4 $ GHz and $ B=100 $ MHz, respectively, and the noise power is $ \sigma^{2}=3.98 \times 10^{-15} $ W. Upon denoting the distance between BS and WD, BS and IRS, as well as IRS and WD by $ d_{\rm{BW}} $, $ d_{\rm{BI}} $ and $ d_{\rm{IW}} $, respectively, the distance-dependent large-scale fading model is given by
		\begin{equation}\label{Eq68}
			L\left( d \right)=C_{0}\left( \frac{d}{d_{0}} \right) ^{-\kappa}\!, \quad d \in \left\lbrace d_{BW}, d_{BI}, d_{IW}   \right\rbrace,
		\end{equation}	
		where $ \kappa $ represents the path loss exponent, and $ C_{0} $ denotes the path loss at the reference distance $d_{0}=1 $ m. Here, we set $ C_{0}=-30 $ dB. Besides, we assume that the path loss exponent of the BS-WD link, BS-IRS link, and IRS-WD link are $ \kappa_{\rm{BW}}=3.5 $, $ \kappa_{\rm{BI}}=2.2 $ and $ \kappa_{\rm{IW}}=2.2 $, respectively \cite{ref24}. For the small-scale fading, we consider a Rice fading channel model, thus the channel $ \mathbf{H} $ of BS-WD can be expressed as
		\begin{equation}
			\mathbf{H}=\sqrt{\frac{\beta_{\mathrm{BW}}}{1+\beta_{\mathrm{BW}}}} \mathbf{H}^{\mathrm{LoS}}+\sqrt{\frac{1}{1+\beta_{\mathrm{BW}}}} \mathbf{H}^{\mathrm{NLoS}},
		\end{equation}
		where $ \beta_{\mathrm{BW}} $ is the Rician factor, $ \mathbf{H}^{\mathrm{LoS}} $ and $ \mathbf{H}^{\mathrm{NLoS}} $ represent the LoS and non-LoS Rayleigh fading components, respectively. Note that $ \mathbf{H} $ is equivalent to a LoS channel when $ \beta_{\mathrm{BW}} \rightarrow \infty $, and a Rayleigh fading channel when $ \beta_{\mathrm{BW}}=0$. Here, let $\beta_{\mathrm{BW}}$, $\beta_{\mathrm{BI}}$, and $\beta_{\mathrm{IW}}$ denote the Rician factors of the BS-WD link, BS-IRS link and IRS-WD link, respectively. We further set $ \beta_{\mathrm{BW}}=0  $, $ \beta_{\mathrm{BI}} \rightarrow \infty  $ and $ \beta_{\mathrm{IW}}=0  $ as \cite{ref7}. The complete BS-WD channel can be obtained by multiplying the small scale fading $ \mathbf{H} $ by the square root of the large scale path loss $L\left( d \right)$. Likewise, the BS-IRS channel and the IRS-WD channel can also be generated through the aforementioned procedure. The default setting of “Communication setting”  are presented in Table \ref{tab2}. In addition, the computation-related settings are provided in the “Computing setting” block of Table \ref{tab2} and the value of these parameter are referenced from \cite{ref24}.
		
		\begin{figure*}[htbp]
			\small
			\captionsetup{font={small}}
			\centering
			\subfloat[Convergence of Algorithm \ref{alg4}]{\includegraphics[width=0.3\textwidth]{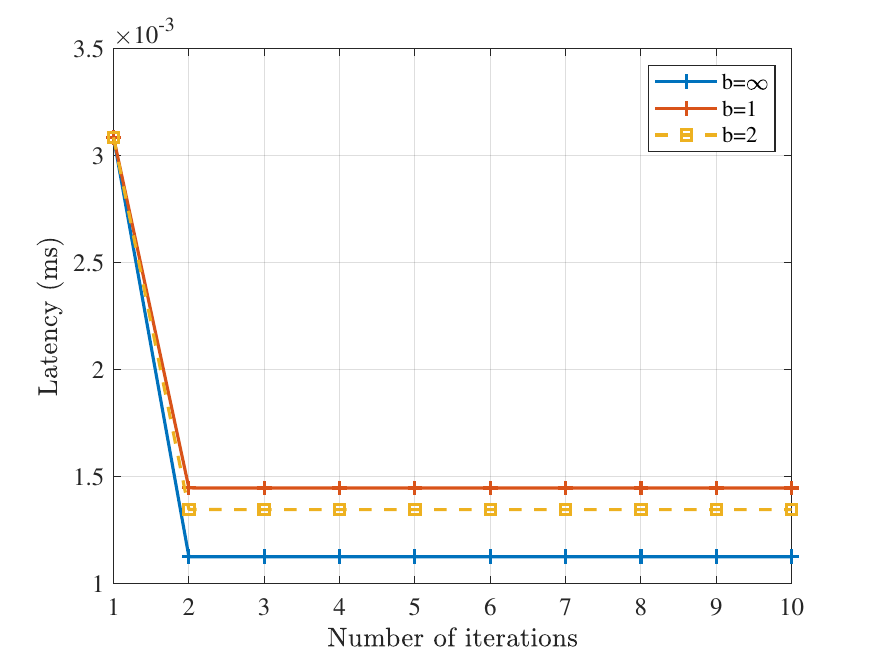}}\label{fig2.1}
			\hfill
			\subfloat[Convergence of Algorithm \ref{alg2}]{\includegraphics[width=0.3\textwidth]{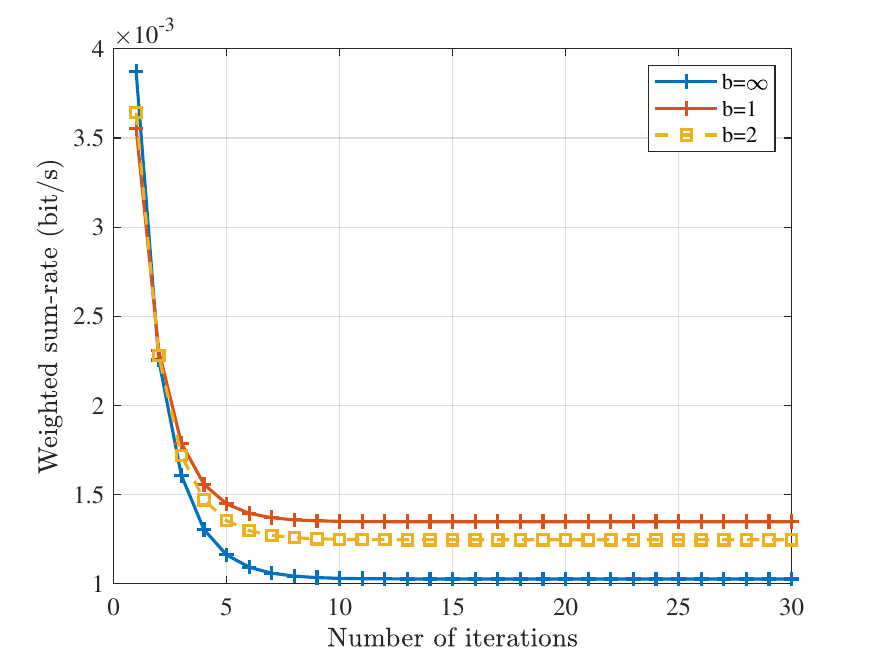}}\label{fig2.2}
			\hfill
			\subfloat[Convergence of Algorithm \ref{alg3}]{\includegraphics[width=0.3\textwidth]{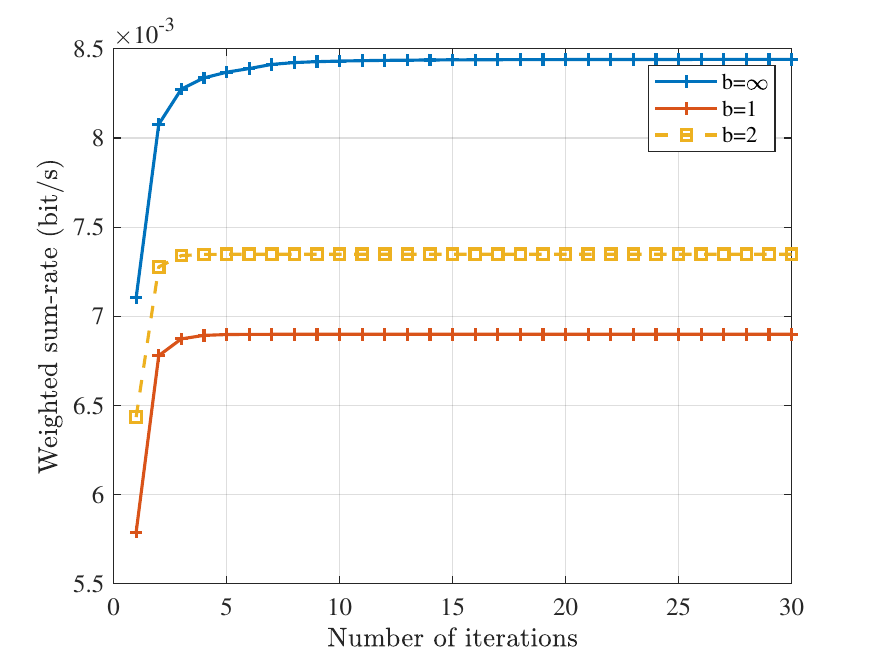}}\label{fig2.3}
			\caption{Convergence hehavior of the proposed algorithm} 
			\label{fig2}
		\end{figure*}
\begin{figure*}[htbp]
			\begin{minipage}[t]{0.42\textwidth}
				\centering
				\includegraphics[width=1.2\textwidth]{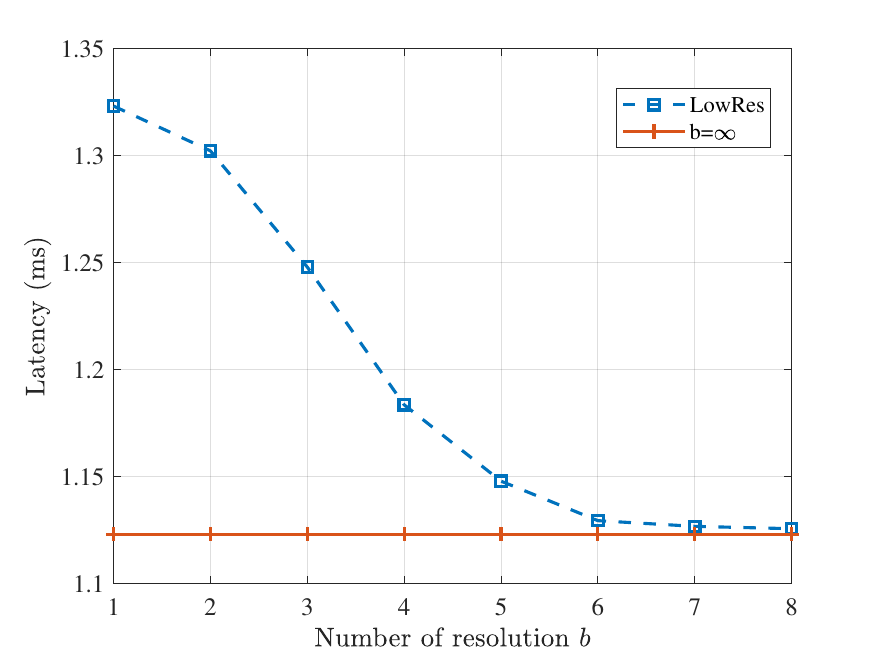}
				\caption{Latency versus the resolution $b$.}
				\label{fig3}
			\end{minipage}
			\qquad \qquad
			\begin{minipage}[t]{0.42\textwidth}
				\centering
				\includegraphics[width=1.2\textwidth]{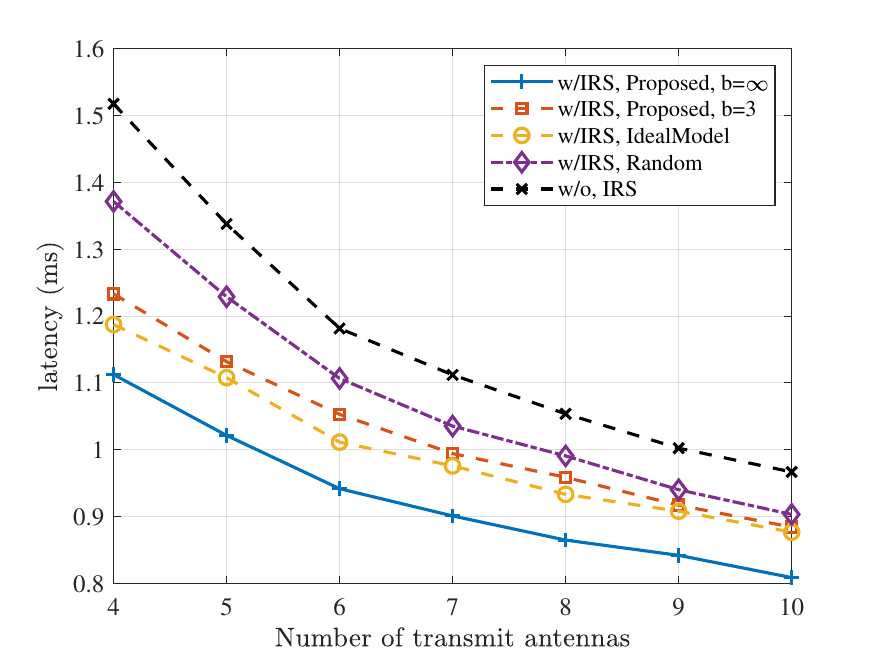}
				\caption{Latency versus the number of transmit antennas $M$.}
				\label{fig4}
			\end{minipage}
		\end{figure*}

		In Fig. \ref{fig2}, we first demonstrate the convergence of the proposed algorithm by depicting latency versus the number of iterations. Numerical results show that Algorithm \ref{alg4} and Algorithm \ref{alg2} can achieve convergence within 2 iterations and 10 iterations, respectively, regardless of using continuous phase shifters or low-resolution phase shifters. Additionally, simulation results show that Algorithm \ref{alg3} can achieve convergence within 10 iterations for continuous phases case and within 5 iterations for low-resolution phases case. Fig. \ref{fig3} depicts the latency versus resolution $b$. It is observed that a resolution level of $b=5$ is deemed adequate, with only marginal performance enhancements when the value of $b$ is greater than 5. Examing both the convergence behavior illustrated in Fig. \ref{fig2} (c) and the impact of resolution $b$ presented in Fig. \ref{fig3}, it is more cost-effective to apply low-resolution phase shifters at IRSs.
		
		Next, the performance of the proposed algorithm is evaluated for the continuous phases case and discrete phases case (i.e., $b=3  $ bits) under different system settings. The following three benchmark schemes are taken into consideration for a fair comparison.
		\begin{itemize}
			\item \emph{w/IRS, Proposed:} The latency optimization by our proposed algorithm based on the practical IRS reflecting model. 
			\item \emph{w/IRS, Ideal: } The IRS optimization design based on the ideal reflection model.
			\item \emph{w/o, IRS:} Traditional MEC systems without IRS.
			\item \emph{w/IRS, Random:} Set the value of BPS $\boldsymbol{\theta}$ randomly within the range $\left[-\pi, \pi\right] $.
		\end{itemize}
		
		Fig. \ref{fig4} shows the latency versus the number of BS's transmit antennas. The “w/o, IRS” scheme achieves the highest latency, illustrating the benefits of introducing IRS into MEC systems. The latency gap between the “w/IRS, Ideal” and the “w/IRS, Proposed” schemes demonstrates the significance of considering the non-ideal IRS reflecting model.
		
\begin{figure*}[htbp]
			\begin{minipage}[t]{0.42\textwidth}
				\centering
				\includegraphics[width=1.2\textwidth]{fig5.pdf}
				\caption{Latency versus the number of IRS elements $N$.}
				\label{fig5}
			\end{minipage}
			\qquad \qquad
			\begin{minipage}[t]{0.42\textwidth}
				\centering
				\includegraphics[width=1.2\textwidth]{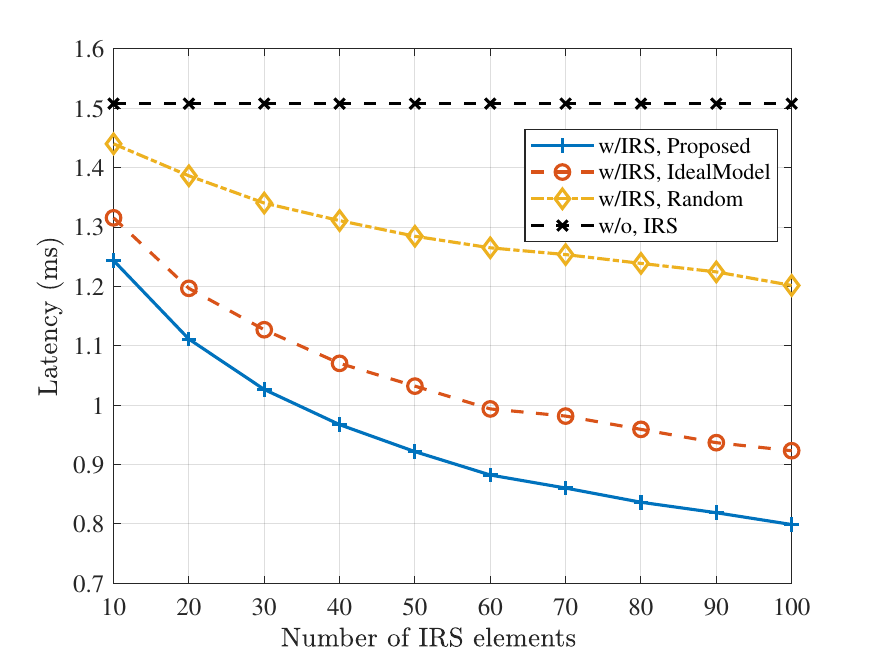}
				\caption{Latency versus the edge computing capability $F_{total}^{e}$.}
				\label{fig6}
			\end{minipage}
		\end{figure*}

		The latency as a function of the number of IRS elements is presented in Fig. \ref{fig5}. The same conclusion can be drawn from Fig. \ref{fig5} that the proposed algorithm outperforms its competitors in term of latency performance. Moreover, the performance gap between the schemes “w/IRS, Proposed” and the “w/IRS, Ideal” becomes higher upon increasing the number of IRS elements. The reason is that the signal power received by BS from the direct link dominates when $N$ is of a small value, whereas the power received from the WD-IRS-BS link becomes dominant as the value of $N$ increases. Consequently, the performance degradation resulting from hardware circuit defects of IRS becomes more significant.
		
		Fig. \ref{fig6} illustrates the latency as a function of the edge computing capability $ F_{\rm{total}} $. It is evident that, for all schemes, the latency drastically reduces upon increasing $F_{\rm{total}}$ when the value of $F_{\rm{total}}$ is small, while the rate of latency reduction diminishes as $F_{\rm{total}}$ approaches a specific threshold, specifically  $30 \times 10^{11}$ cycle/s. The reason for this phenomenon is that the edge computing latency dominates when the value of $F_{\rm{total}}$ is small, whereas the data transmission latency becomes dominant when $F_{\rm{total}}$ reaches a sufficiently large level. This indicates that adding appropriate computing capability to edge server is a more economical method for latency minimization. 
		
\begin{figure*}[htbp]
			\begin{minipage}[t]{0.42\textwidth}
				\centering
				\includegraphics[width=1.2\textwidth]{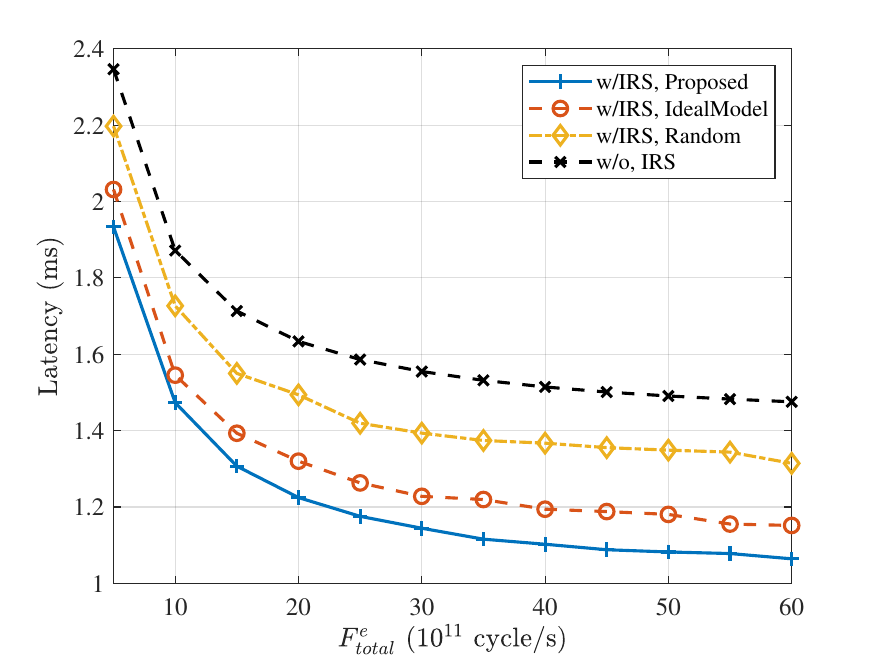}
				\caption{Latency versus the distance $ L $.}
				\label{fig7}
			\end{minipage}
			\qquad \qquad
			\begin{minipage}[t]{0.42\textwidth}
				\centering
				\includegraphics[width=1.2\textwidth]{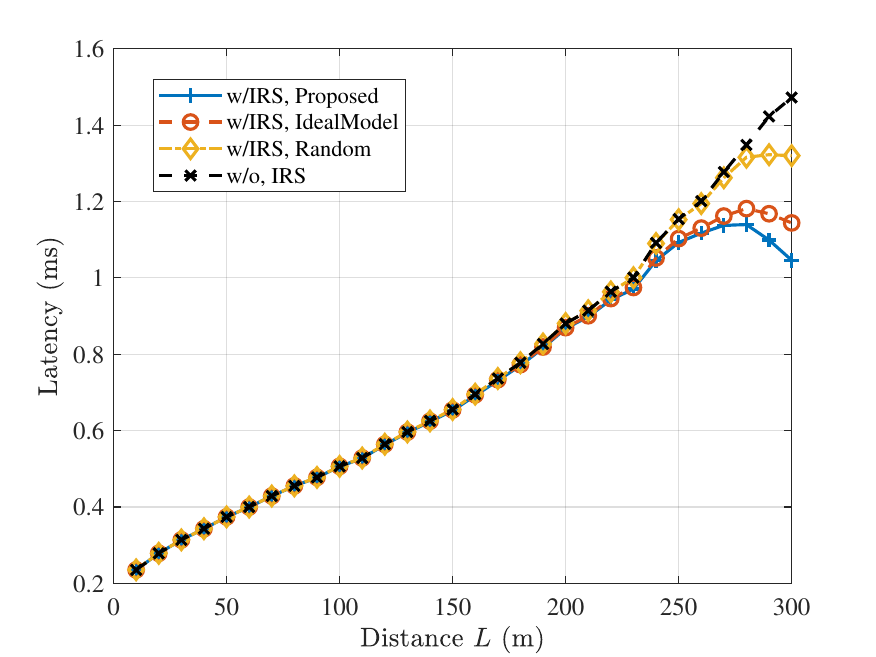}
				\caption{Latency versus the number of users $ K $.}
				\label{fig8}
			\end{minipage}
		\end{figure*}

		Fig. \ref{fig7} shows the latency versus the distance $L$. Our observations are as follows. Firstly, for the conventional scheme without IRS, the latency increases upon increasing the distance $L$. Secondly, when WDs move away from BS to the IRS, the latency increases at first and reaches its maximum value at $d=280$~ m, because the signal power received by BS from WDs becomes weaker, and then decreases thanks to the reflecting signal from IRS. Thirdly, for the “w/IRS, Proposed” scheme, the advantages of employing IRS become notable when the distance between WD and IRS is less than 100 m, whereas the beneficial role of IRS becomes visible only when the distance is less than 20 m for the “w/IRS, Random” scheme. It is important to note that our proposed algorithm can achieve better latency performance compared with others under different distance settings.
		
		Finally, Fig. \ref{fig8} illustrates the latency as a function of the number of WDs. It is evident that the average latency rises as the number of WDs increases. This can be attributed to the diminished beamforming gain achieved at each WD and the reduced edge computing resources allocated to each WD. The former problem may be relieved by deploying more IRS to form stronger beams, and the latter issue can be solved by enhancing the computing capabilities of edge servers. Moreover, when $K=5$, the proposed algorithm can reduce latency from 2.24 ms to 2.07 ms, compared to the “w/IRS, Ideal” scheme. This further validates the benefits of the proposed algorithm.
		
		\section{Conclusion}
		To maximize the benefits of MEC systems, this paper studied a weighted latency minimization problem in an IRS-enhanced wideband MEC system with non-ideal reflection model. Both the limited edge computing resources and non-ideal IRS model are considered. To deal with the formulated non-convex problem, we used the BCD technique to decouple it into computing design subproblem and communication design subproblem, which were then optimized alternately. Simulation results confirmed the effectiveness of the proposed algorithm and the importance of utilizing the non-ideal reflection model for designing IRS-enhanced wideband MEC systems. Furthermore, numerical results demonstrated the rapid convergence of our proposed algorithm.
		
		Finally, it is important to note that this paper is based on the hypothesis of perfect channel state information (CSI) for all the related communication links which may not be realistic for practical systems. Therefore, considering the design of IRS-enhanced MEC systems under imperfect CSI is left as one of our future works.

	\end{document}